\documentclass[preprint,aps,prc,nofootinbib,superscriptaddress,showpacs,showkeys]{revtex4}
\usepackage{epsfig}
\usepackage{graphicx}
\usepackage{amsfonts}
\usepackage{amsmath}
\begin{document}

\title{Numerical values of $f^F$, $f^D$, $f^S$ coupling constants in
$SU(3)$ invariant interaction Lagrangian of vector-meson nonet
with $1/2^+$ octet baryons}
\date{\today}

\author{Cyril Adamuscin}
\address{Institute of Physics, Slovak Academy of Sciences,
        Bratislava}
\author{Erik Bartos}
\address{Institute of Physics, Slovak Academy of Sciences,
        Bratislava}
\author{Stanislav Dubnicka}
\address{Institute of Physics, Slovak Academy of Sciences,
        Bratislava}
\author{Anna Zuzana Dubnickova}
\address{Department of Theoretical Physics, Comenius University,
        Bratislava}

\begin{abstract}

   It is demonstrated how an utilization of all existing
experimental information on electric and magnetic nucleon form
factors, to be described by the Unitary and Analytic (U\&A)
nucleon electromagnetic structure model in space-like and
time-like regions simultaneously, can provide numerical values of
$f^F$, $f^D$, $f^S$ coupling constants in $SU(3)$ invariant
interaction Lagrangian of the vector-meson nonet with $1/2^+$
octet baryons. The latter, together with universal vector-meson
coupling constants $f_V$, play an essential role in a prediction
of $1/2^+$ octet hyperon electromagnetic form factors behaviors.
\end{abstract}



\maketitle

\section{Introduction}

   Recently the BESIII Collaboration has published \cite{ablikim} new
results on the proton electromagnetic (EM) form factors (FFs) in
the time-like region by measuring the process $e^+e^- \to p \bar
p$ with very high precision at 12 different center-of-mass
energies from $4.9836$ GeV$^2$ to $13.4762$ GeV$^2$. In such way
the obtained results enriched a set of already existing number of
experimental points on proton EM FFs in the space-like and
time-like regions substantially. Moreover, the BESIII Collab. and
before also the BaBar Collab. \cite{lees}, measuring the polar
angular distribution $F(\cos \theta_p)$ of created protons at few
different energies, have been able to determine a separate
information on the proton electric and on the proton magnetic FFs
in time-like region for the first time.

On the other hand a very successful microscopic model for the proton EM FFs
in the near-threshold time-like region has been elaborated \cite{xi} to be based
on the assumption that the behavior of the EM FFs as a function of the energy is
given by the initial or final state interaction between proton
and antiproton in the processes $e^+e^- \leftrightarrow p \bar p$.
In this approach the antinucleon-nucleon potential is constructed in the
framework of chiral effective field theory \cite{xiwu} and fitted to
results of partial-wave analysis of existing $\bar p p$ scattering data. As a
result predictions of the proton EM FFs $G_E^p(t)$  and $G_M^p(t)$, and also
their ratios are in good agreement with existing data up to $t=3.986\; {\rm GeV}^2$,
i.e. at the region of a validity of the model under consideration.

   Next intentions of the BESIII Collab. are to extend such data
measurements for $e^+e^- \to Y \bar Y$ processes, where $Y$ are
hyperons to be members of the $1/2^+$ octet baryons

   $p$, $n$, $\Lambda$, $\Sigma^+$, $\Sigma^0$, $\Sigma^-$, $\Xi^0$,
   $\Xi^-$.

   In a preparation of such a program it can be useful to dispose
with some model predictions for hyperon EM FF behaviors and
subsequently also of the corresponding differential and total
cross-sections.

But the model presented in \cite{xi} is not applicable for such model predictions of
hyperon EM FFs as it is based on the existence of baryon-antibaryon scattering data,
what is unattainable in the case of hyperons.

   In this paper we demonstrate a scheme in the framework of which
one can do such predictions. As one can see further, an important
role in the prediction of hyperon EM FFs play the advanced Unitary
and Analytic ($U\&A$) nucleon EM structure model, respecting the
$SU(3)$ symmetry and OZI rule \cite{okubo}-\cite{iizuka}
violation, then a knowledge of the universal vector-meson coupling
constants $f_V$ to be extracted from the experimental values of
vector-meson lepton widths

\begin{equation}
  \Gamma(V \to e^+e^-)=\frac{\alpha^2
  m_V}{3}\bigg(\frac{f_V^2}{4\pi}\bigg)^{-1}\label{lwidth}
\end{equation}
with $\alpha=\frac{1}{137}$ to be the QED fine structure coupling
constant and $m_V$ the vector-meson mass, and a knowledge of the
numerical values of $f^F$, $f^D$, $f^S$ coupling constants in $SU(3)$
invariant Lagrangian

\begin{eqnarray}
L_{VB\bar{B}}=\frac{i}{\sqrt{2}}f^F[\bar{B}^\alpha_\beta
\gamma_\mu B^\beta_\gamma- \bar{B}^\beta_\gamma\gamma_\mu
B^\alpha_\beta](V_\mu)^\gamma_\alpha+\\\nonumber
+\frac{i}{\sqrt{2}}f^D[\bar{B}^\beta_\gamma \gamma_\mu
B^\alpha_\beta+
\bar{B}^\alpha_\gamma\gamma_\mu B^\beta_\gamma](V_\mu)^\gamma_\alpha+\\ \nonumber
+\frac{i}{\sqrt{2}}f^S \bar{B}^\alpha_\beta \gamma_\mu
B^\beta_\alpha\omega^0_\mu\nonumber \label{lagrangian}
\end{eqnarray}
describing strong interactions of the nonet of vector-mesons with
$1/2^+$ octet baryons, where $B, \bar B$ and $V$ are baryon,
anti-baryon and vector-meson octuplet matrices and $\omega ^0_\mu$
is omega-meson singlet.

   The advanced $U\&A$ nucleon EM structure model represents a
harmonious unification of the vector-meson pole contributions and
cut structure of EM FFs, which represent the so-called continua
contributions in nucleon EM FFs.

   The nucleon EM FFs are analytic functions in the whole complex
plane of their variable $t$, besides cuts on the positive real
axis from the lowest branch point $t_0$ to $+\infty$.

   The shape of nucleon EM FFs is directly related with an
existence of complex conjugate pairs of poles on unphysical sheets
of the Riemann surface in $t$-variable, corresponding to unstable
true neutral vector-mesons \cite{olive} $\rho(770)$, $\omega(782)$,
$\phi(1020)$; $\rho'(1450)$, $\omega'(1420)$, $\phi'(1680)$;
$\rho''(1700)$, $\omega''(1650)$, $\phi''(2170)$ with the quantum
numbers of the photon to be revealed explicitly in the process of
electron-positron annihilation  into hadrons.

   As a result the complex nature of nucleon EM FFs for $t>t_0$ in
time-like region is secured by imaginary parts of the vector-meson
poles on unphysical sheets of the Riemann surface in $t$-variable
and the cuts on the positive real axis of the first, so-called
physical sheet, of the Riemann surface, whereby imaginary part of
the EM FFs are given by a difference of the FF values on the upper
boundary of the cuts and the FF values on the lower boundary of
the cuts.

   Every nucleon electric FF is canonically normalized to the
nucleon electric charge and every nucleon magnetic FF is
normalized to the nucleon magnetic moment.

   All these nucleon EM FFs govern the asymptotic behaviors as
predicted by the quark model of hadrons to be proven also in the
framework of the QCD \cite{brodskylepage}.

   The advanced $U\&A$ nucleon EM structure model depends on the
coupling constants ratios $(f_{VNN}/f_V)$ to be defined for every
of nine above-mentioned unstable true neutral vector-mesons in VMD
model. However, not all nine of them appear explicitly in the
advanced $U\&A$ nucleon EM structure model as free parameters. In
the process of a construction of the VMD model to be automatically
normalized and governing the correct asymptotic behavior, some of
these coupling constants ratios can be expressed through the table
mass values of the true neutral vector-mesons under consideration
and other coupling constant ratios. And just here some freedom
exists in the choice, which of the coupling constants ratios
$(f_{VNN}/f_V)$ will be left as free parameters of the model.

   The problem is that not for all nine true neutral vector-mesons
under consideration an experimental information on the lepton
width $\Gamma(V \to e^+e^-)$ exists. The latter is known
\cite{olive} only for ground state vector-mesons $\rho(770)$,
$\omega(782)$ and $\phi(1020)$. For all the first and second
excited states the Rev. of Part. Physics \cite{olive} declares,
that their lepton decays are seen, but they are not precisely
measured experimentally up to now.

   Just for this reason in the present advanced nucleon EM
structure model we keep the coupling constant ratios $(f_{\rho
NN}/f_\rho)$, $(f_{\omega NN}/f_\omega)$, $(f_{\phi NN}/f_\phi)$ and
also $(f_{\omega' NN}/f_{\omega'})$, $(f_{\phi' NN}/f_{\phi'})$, for
which some model estimations of $f_{\omega'}$ and $f_{\phi'}$
exist \cite{donnachieclegg}, as free parameters to be determined
in comparison of the present advanced nucleon EM structure model
with all existing data on nucleon EM FFs in space-like and
time-like regions simultaneously.

   The estimation of $f_{\omega'}$, $f_{\phi'}$ (and also $f_{\rho'}$ for
a determination of $f_{\rho'NN}$ from ($f_{\rho'NN}/f_{\rho'}$) to
be completely given by the table mass values of vector-mesons
under consideration) in \cite{donnachieclegg} will be only a model
admixture in subsequent determination of the $1/2^+$ octet hyperon
EM FF behaviors.

\section{Existing experimental information on nucleon EM FFs}

   The EM structure of the nucleons (iso-doublet compound of the
proton and neutron), as revealed experimentally for the first time
in elastic unpolarized electron-proton scattering in the middle of
the last century, is completely described by four independent
scalar functions, the electric $G_{Ep}(t)$,$ G_{En}(t)$ and the
magnetic $G_{Mp}(t)$, $G_{Mn}(t)$ FFs, dependent of one variable to
be chosen as the squared momentum transferred $t=-Q^2$ of the
virtual photon $\gamma^*$. The experimental information on these
functions consists of 11 following different sets of data on:
\begin{itemize}
\item
  the ratio $\mu_p G^p_E(t)/G^p_M(t)$ in the space-like $(t<0)$ region
  from polarization experiments \cite{jones}-\cite{puckett}
\item
  $G^p_E(t)$ in the space-like $(t<0)$ region \cite{bernauer}
\item
  $|G^p_E(t)|$ in the time-like $t>0$ region; only from exp. in
  which $|G^p_E(t)|$=$|G^p_M(t)|$ is assumed
\item
  $G^p_Mt)$ in the space-like $(t<0)$ region
  \cite{bernauer}-\cite{andivakis}
\item
  $|G^p_M(t)|$ in the time-like $t>0$ region
  \cite{ablikim},\cite{lees},\cite{bassompierre}-\cite{aubert}
\item
  $|G^p_E(t)/G^p_M(t)|$ in the time-like $t>0$ region
  \cite{ablikim},\cite{lees}
\item
  $G^n_E(t)$ in the space-like $(t<0)$ region \cite{hanson}-\cite{golak}
\item
  $|G^n_E(t)|$ in the time-like $t>0$ region; only from exp. in
  which $|G^n_E(t)|$=$|G^n_M(t)|$ is assumed
\item
  $G^n_M(t)$ in the space-like $(t<0)$ region
  \cite{hanson},\cite{rock}-\cite{anderson}
\item
  $|G^n_M(t)|$ in the time-like $t>0$ region \cite{antonelli}
\item
  the ratio $\mu_n G^n_E(t)/G^n_M(t)$ in the space-like $(t<0)$ region
  from polarization experiments on the light nuclei
  \cite{plaster},\cite{riordan}
\end{itemize}
from which not all are equally trustworthy.

   The most reliable, from all above-mentioned data, are
considered to be the experimental points on the ratio $\mu_p
G^p_E(t)/G^p_M(t)$ \cite{jones}-\cite{puckett} in the space-like
region, which have been extracted from simultaneous measurement of
the transverse component

\begin{equation}
  P_t=\frac{h}{I_0}(-2)\sqrt{\tau(1+\tau)}G_{Ep}G_{Mp}\tan{\theta/2}\label{transpol}
\end{equation}
and the longitudinal component
\begin{equation}
  P_l=\frac{h(E_e+E_{e'})}{I_0m_p}\sqrt{\tau(1+\tau)}G_{Mp}^2\tan^2{\theta/2}\label{longpol}
\end{equation}
of the recoil proton's polarization in the polarized electron
scattering plane of the polarization transfer process
$\overrightarrow{e}p \to e\overrightarrow{p}$, where $h$ is the
electron beam helicity, $I_0$ is the unpolarized cross-section
excluding $\sigma_{Mott}$ and $\tau=-\frac{t}{4m^2_p}$, by means
of the relation

\begin{equation}
  \mu_p\frac{G_{Ep}}{G_{Mp}}=-\frac{P_t}{P_l}\frac{(E_e+E_{e'})}{2m_p}\tan{\theta/2}\label{ratiosl}.
\end{equation}

   The data \cite{jones}-\cite{puckett} clearly demonstrate that a
general belief in the dipole behavior of the proton electric FF
$G^p_E(t)$ in the space-like region to be obtained by the
Rosenbluth method  from the process of the elastic scattering of
unpolarized electrons on unpolarized protons $e^-p \to e^-p$
described by the differential cross-section in the laboratory
system
\begin{eqnarray}
 \frac{d\sigma^{lab}(e^- p \to e^- p)}{d\Omega}=\frac{\alpha^2}{4E^2}\frac{\cos^2(\theta/2)}{\sin^4(\theta/2)}
 \frac{1}{1+(\frac{2E}{m_B})\sin^2(\theta/2)}\label{diffcsec}\\
.\bigg[\frac{G_E^2(t)-\frac{t}{4m_B^2}G_M^2(t)}{1-\frac{t}{4m_B^2}}-2\frac{t}{4m_B^2}G_M^2(t)
\tan^2(\theta/2)\bigg]\nonumber
\end{eqnarray}
with the QED the fine structure constant $\alpha=1/137$, the
incident electron energy $E$ and the scattering angle $\theta$,
before the polarization experiments \cite{jones}-\cite{puckett}
have been carried out, is no more valid.

   This is a reason, why further we ignore all older data on
$G^p_E(t)$ in the space-like region and we take into account only
the newest data from MAMI \cite{bernauer} at the region
$-0.5524GeV^2<t<-0.0152GeV^2$, where the application of the
Rosenbluth method is still justified.

   On the second place to be concerned of the reliability are the
data on the proton magnetic FF $G^p_M(t)$
 \cite{bernauer}-\cite{andivakis} in the space-like
region, which have been obtained by the Rosenbluth method from
experimental information on the differential cross-section
describing elastic scattering of unpolarized electrons on
unpolarized protons. As one can see from the expression of the
differential cross-section (\ref{diffcsec}), with increased negative values of
$t$, the proton magnetic FF is dominant in comparison with the
proton electric FF and so, the data on $G^p_M(t)$ extracted by the
Rosenbluth method are reliable.

   Less reliable are data \cite{ablikim},\cite{lees},\cite{bassompierre}-\cite{aubert}
on absolute value of the magnetic FF $|G^p_M(t)|$ in the time-like
$t>0$ region (the same is concerned also of the neutron magnetic
FF $|G^n_M(t)|$ in the time-like $t>0$ region) as the most of them
are extracted from the total cross-section of the
electron-positron annihilation process into proton-antiproton pair
\begin{equation}
\sigma_{tot}^{c.m.}(e^+ e^- \to p \bar p)=\frac{4\pi \alpha^2
\beta_p}{3t}[\mid G^p_M(t)\mid^2+\frac{2m_p^2}{t}\mid
G^p_E(t)\mid^2]\label{dirtotcsec}
\end{equation}

with $\beta_p=\sqrt{1-\frac{4m_p^2}{t}}$, or from the total
cross-section of the antiproton-proton annihilation into
electron-positron pair
\begin{equation}
\sigma_{tot}^{c.m.}(\bar p p \to e^+ e^-)=\frac{2\pi
\alpha^2}{3p_{c.m.}\sqrt{t}}[\mid
G^p_M(t)\mid^2+\frac{2m_p^2}{t}\mid
G^p_E(t)\mid^2]\label{invtotcsec}
\end{equation}
with $p_{c.m.}$ to be the antiproton momentum in the c.m. system,
by and assumption $|G^p_E(t)|$=$|G^p_M(t)|$, which is exactly
valid only at the proton-antiproton threshold following directly
from the definition of both FFs, or by an assumption that for high
values of $t$ $|G^p_E(t)|$=0 and then the total cross-sections
under consideration give information only on $|G^p_M(t)|$, what is
not well-founded.

   Very promising method of a determination of $|G^p_E(t)|$ and
$|G^p_M(t)|$ is demonstrated in \cite{ablikim},\cite{lees} by a
measurement of the distribution of $\theta_p$, the angle between
the proton momentum in the $p \bar p$ rest frame , and the
momentum of the $p \bar p$ system in the $e^+e^-$ c.m. frame and
by a subsequent fitting of the obtained experimental points by the
expression

\begin{equation}
   F(\cos\theta_p)=N_{norm}\bigg[1+\cos^2\theta_p+\frac{4m^2_p}{t}R^2(1-\cos^2\theta_p)\bigg]\label{poldistr}
\end{equation}

to be obtained from

\begin{equation}
  \frac{d\sigma(t)}{d\Omega}=\frac{\alpha^2 \beta_p
  C}{4t}\bigg[|G^p_M(t)|^2(1+\cos^2\theta_p)+\frac{4m^2_p}{t}|G^p_E(t)|^2
  \sin^2\theta_p\bigg]\label{eqa}
\end{equation}

with the Coulomb correction factor $C$, where

\begin{equation}
   R=\frac{|G^p_E(t)|}{|G^p_M(t)|}\label{eqb}
\end{equation}
and

\begin{equation}
   N_{norm}=\frac{2\pi \alpha^2 \beta L}{4t}\bigg[1.94+5.04
   \frac{m^2_p}{t}R^2 |G^p_M(t)|^2\bigg]\label{eq3}
\end{equation}
is the overall normalization factor and $L$ is the integrated
luminosity.

   However, the latter method is in the phase of a development and
the ratios $R$ with $|G^p_M(t)|$ at only several values of the
energy $t$ have been determined \cite{ablikim},\cite{lees} up to
now.

   The existing data on $G^n_E(t)$,$G^n_M(t)$ in the space-like
region and the data on $\mu_n G^n_E(t)/G^n_M(t)$ are even more
model dependent as they have been extracted from the scattering
processes of electrons on light nuclei by various theoretical
model ingredients.

\begin{figure}
\includegraphics[scale=0.6]{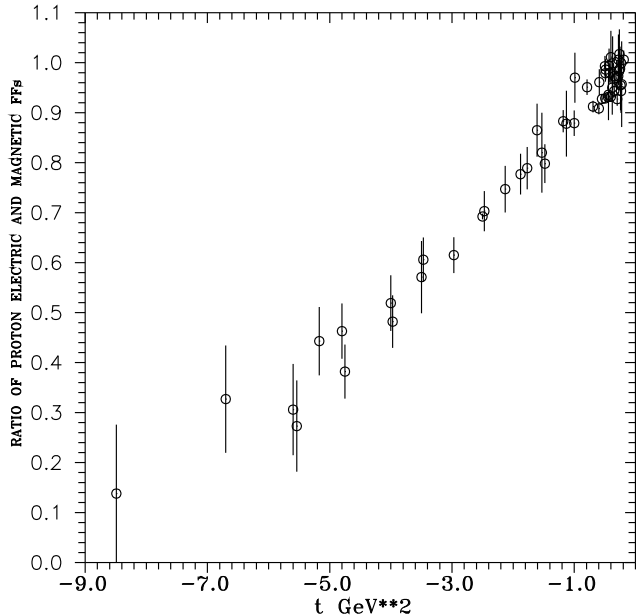}
\caption{Data on the ratio $\mu_pG^p_E(t)/G^p_M(t)$ in the
space-like $(t<0)$ region from polarization experiments
\cite{jones}-\cite{puckett}}
\label{fig1}
\end{figure}

\begin{figure}
\includegraphics[scale=0.6]{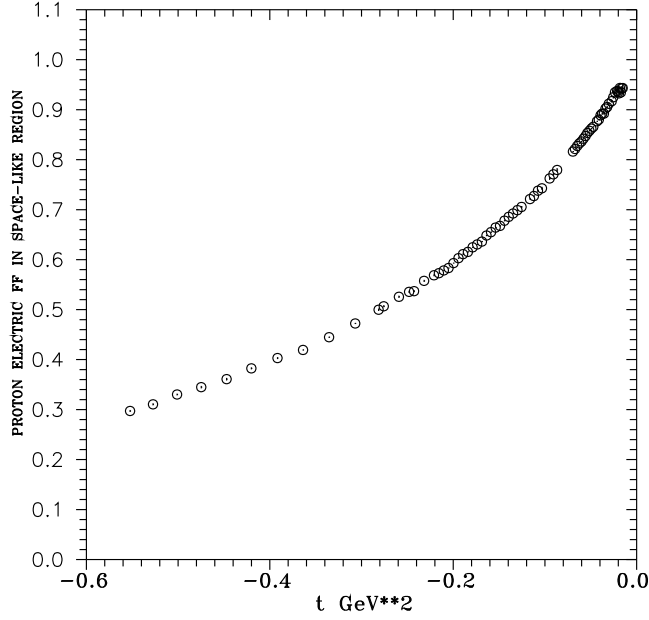}
\caption{Proton electric FF data in the space-like $(t<0)$ region
from MAMI \cite{bernauer}}
\label{fig2}
\end{figure}

\begin{figure}
\includegraphics[scale=0.6]{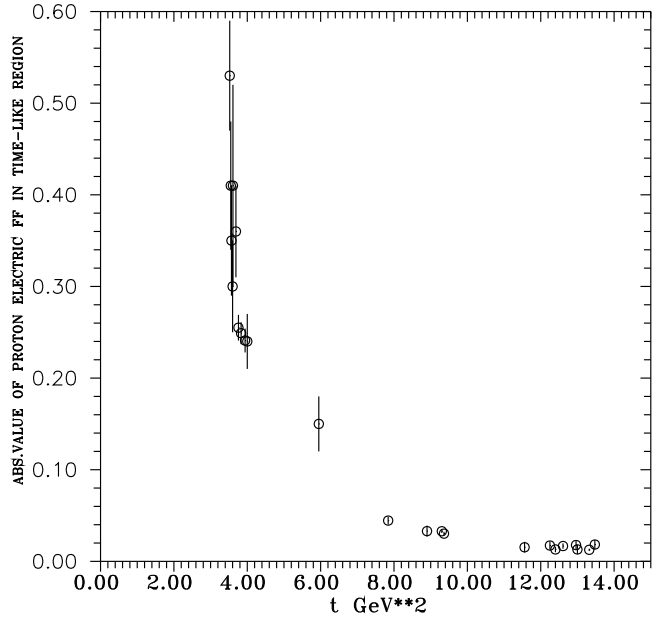}
\caption{Proton electric FF data in the time-like $(t>0)$ region
from exps. in which $|G^p_E(t)|$=$|G^p_M(t)|$ has been assumed}
\label{fig3}
\end{figure}

\begin{figure}
\includegraphics[scale=0.6]{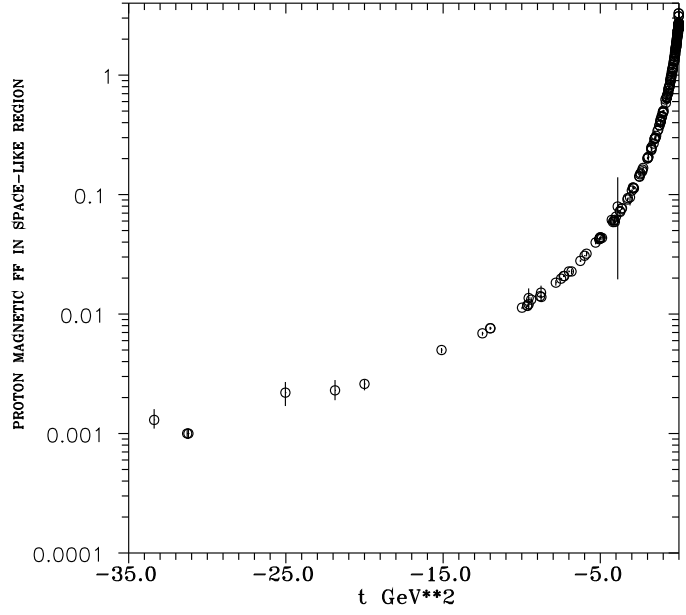}
\caption{Proton magnetic FF data in the space-like $(t<0)$ region
\cite{bernauer}-\cite{andivakis}}
\label{fig4}
\end{figure}

\begin{figure}
\includegraphics[scale=0.6]{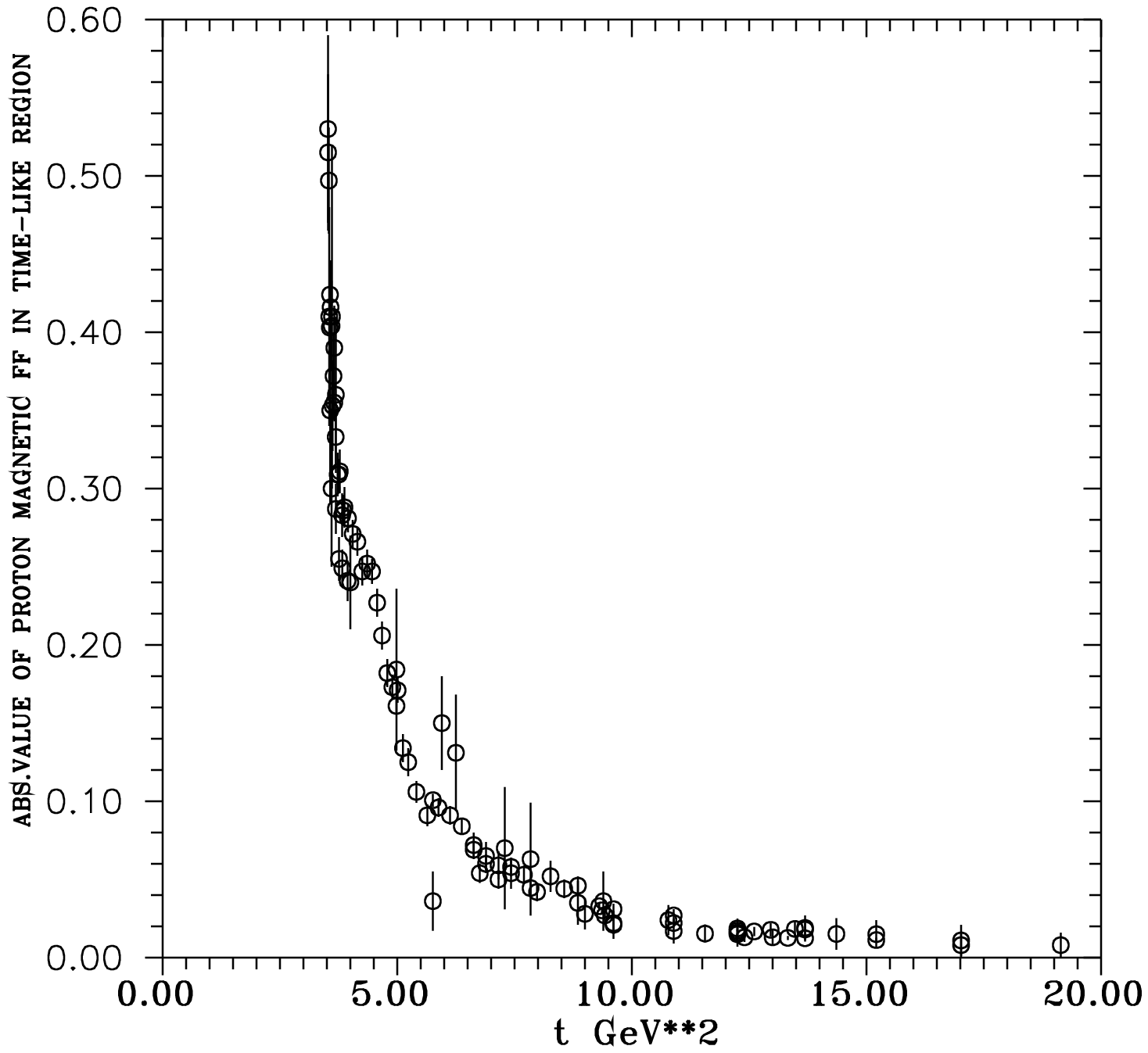}
\caption{Proton magnetic FF data in the time-like $(t>0)$ region
\cite{ablikim},\cite{lees},\cite{bassompierre}-\cite{aubert} }
\label{fig5}
\end{figure}

\begin{figure}
\includegraphics[scale=0.6]{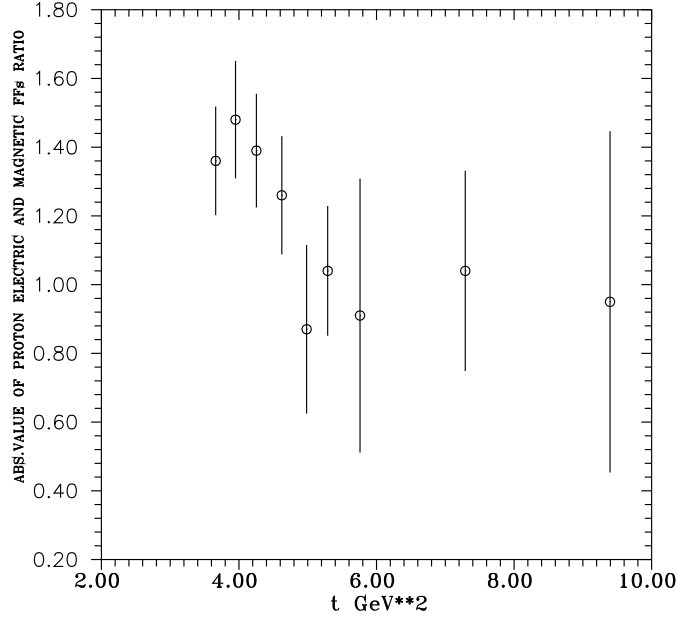}
\caption{Data on the ratio $|G^p_E(t)/G^p_M(t)|$ in the time-like
$(t>0)$ region \cite{ablikim},\cite{lees}}
\label{fig6}
\end{figure}

\begin{figure}
\includegraphics[scale=0.6]{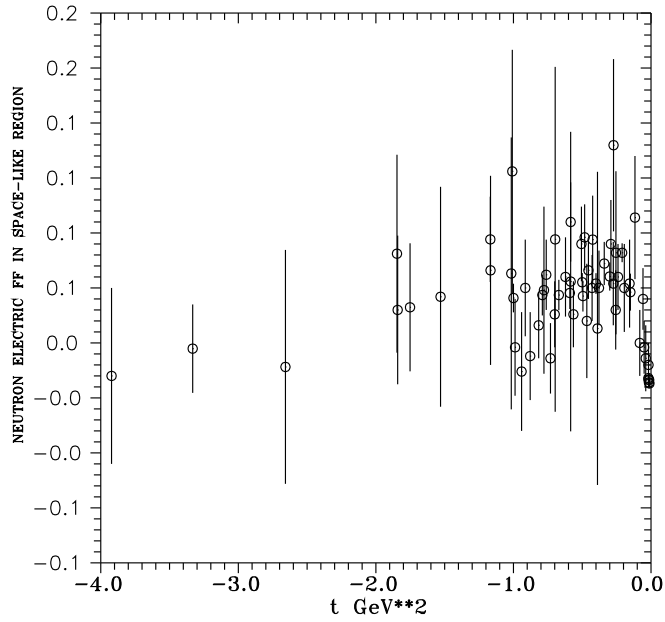}
\caption{Neutron electric FF data in the space-like $(t<0)$ region
\cite{hanson}-\cite{golak}}
\label{fig7}
\end{figure}

\begin{figure}
\includegraphics[scale=0.6]{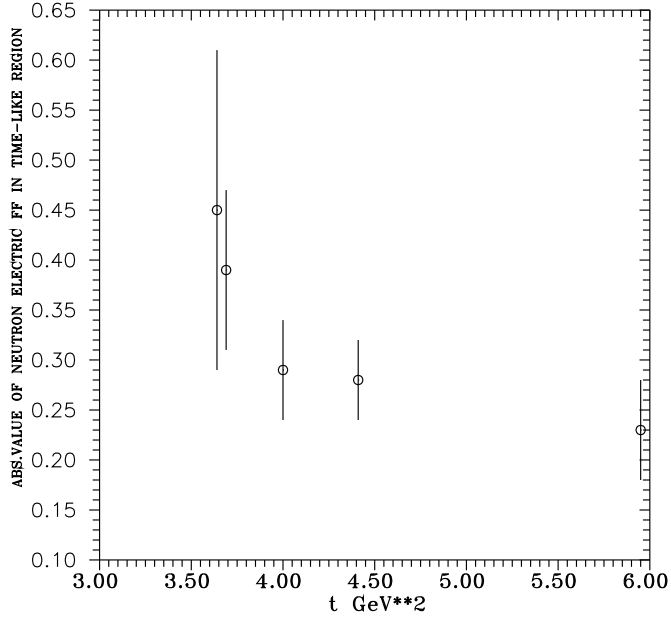}
\caption{Neutron electric FF data in the time-like $(t>0)$ region
from an assumption $|G^n_E(t)|$=$|G^n_M(t)|$}
\label{fig8}
\end{figure}

\begin{figure}
\includegraphics[scale=0.6]{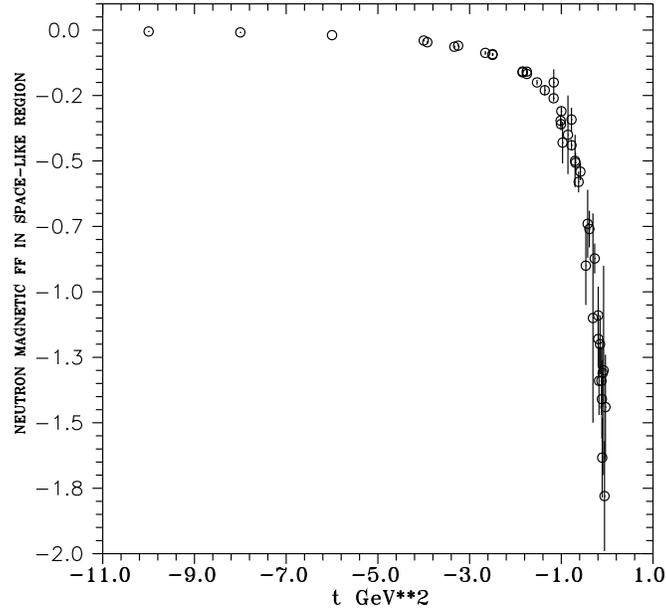}
\caption{Neutron magnetic negative FF data in the space-like
$(t<0)$ region \cite{hanson},\cite{rock}-\cite{anderson}}
\label{fig9}
\end{figure}

\begin{figure}
\includegraphics[scale=0.6]{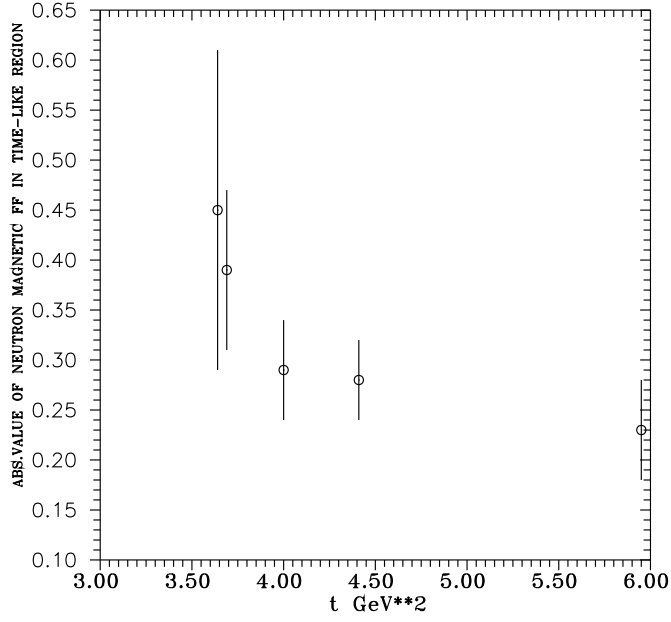}
\caption{Neutron magnetic FF data in the time-like $(t>0)$ region
\cite{antonelli}}
\label{fig10}
\end{figure}

\begin{figure}
\includegraphics[scale=0.6]{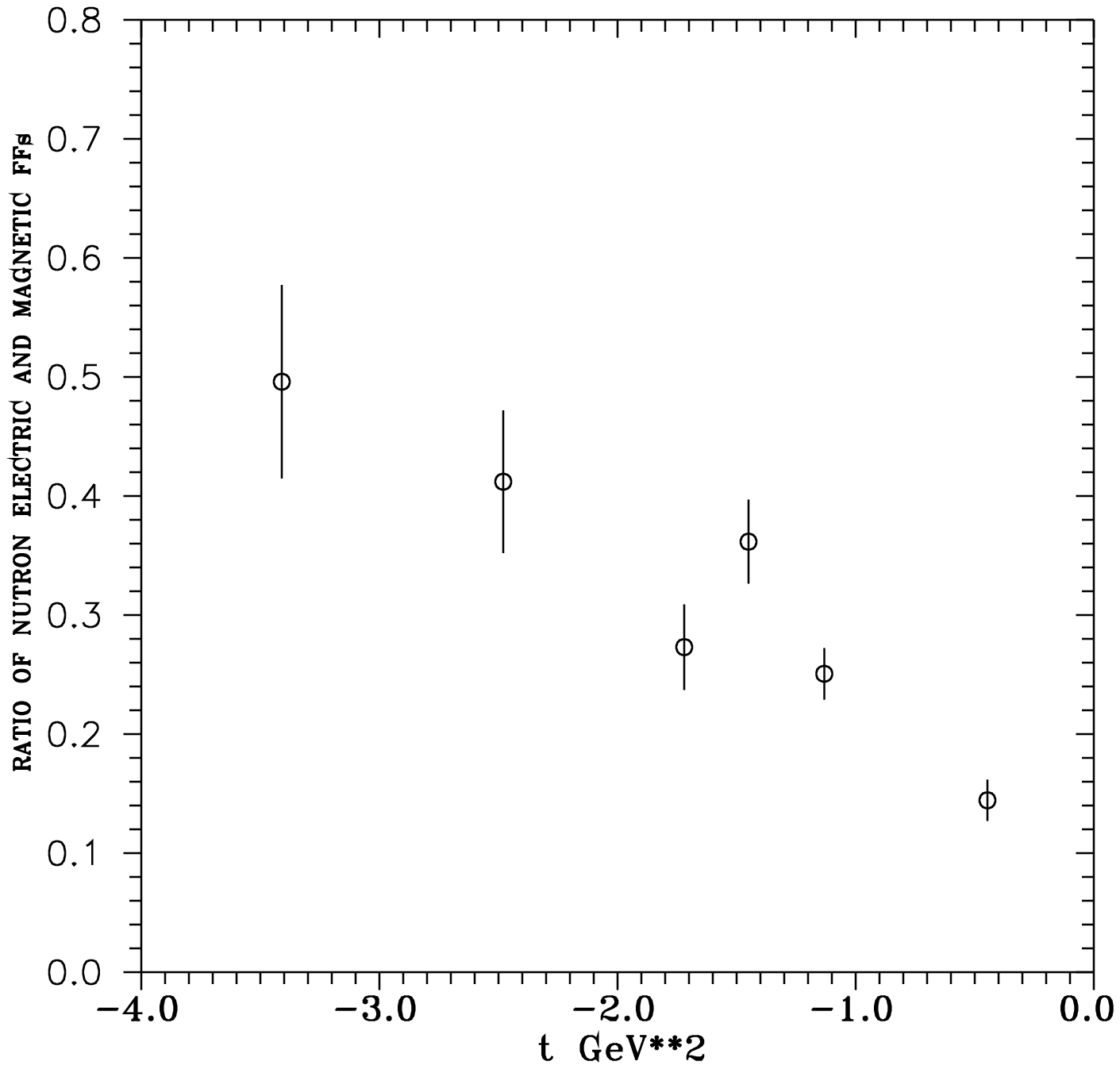}
\caption{Data on the ratio $\mu_nG^n_E(t)/G^n_M(t)$ in the
space-like $(t<0)$ region from polarization experiments on the
light nuclei \cite{plaster},\cite{riordan}}
\label{fig11}
\end{figure}

   All these data are graphically presented in Figs.~(\ref{fig1}-\ref{fig11}).

Despite of all their shortcomings they will be further analyzed by
one advanced $U\&A$ nucleon EM structure model simultaneously.

\section{Advanced Unitary and Analytic nucleon EM structure model}

   First we shall demonstrate, how it is possible in the advanced nucleon
EM structure model to keep the coupling constants ratios $(f_{\rho
NN}/f_\rho), (f_{\omega NN}/f_\omega), (f_{\phi NN}/f_\phi)$ and
also $(f_{\omega' NN}/f_{\omega'}), (f_{\phi' NN}/f_{\phi'})$ as
free parameters of the model to be then subsequently determined
numerically in comparison of the advanced $U\&A$ model with all
existing data simultaneously.

   Nevertheless, before that we would like to note that the
nucleon EM FFs $G^p_E(t), G^p_M(t), G^n_E(t), G^n_M(t)$ are very
suitable for extraction of an experimental information on the
nucleon EM structure from the earlier mentioned physical
quantities (\ref{transpol})-(\ref{poldistr}). But for a
construction of various nucleon EM structure models the
flavour-independent iso-scalar and iso-vector parts $F^N_{1s}(t),
F^N_{1v}(t), F^N_{2s}(t), F^N_{2v}(t)$ of the Dirac and Pauli FFs
to be defined by a parametrization of the matrix element of the
nucleon EM current

\begin{equation}
  <N|J^{EM}_\mu|N>=e\bar u(p')\{\gamma_\mu
  F^N_1(t)+\frac{i}{2m_N}\sigma_{\mu\nu}(p'-p)_\nu
  F^N_2(t)\}u(p)\label{eq4}
\end{equation}
are more suitable.

   Both sets of these FFs are related as follows
\begin{eqnarray}
 G^p_E(t)&=&[F^N_{1s}(t)+F^N_{1v}(t)]+\frac{t}{4m^2_p}[F^N_{2s}(t)+F^N_{2v}(t)]\label{ffrelatios}\\
 G^p_M(t)&=&[F^N_{1s}(t)+F^N_{1v}(t)]+[F^N_{2s}(t)+F^N_{2v}(t)]\nonumber \\
 G^n_E(t)&=&[F^N_{1s}(t)-F^N_{1v}(t)]+\frac{t}{4m^2_n}[F^N_{2s}(t)-F^N_{2v}(t)]\nonumber \\
 G^n_M(t)&=&[F^N_{1s}(t)-F^N_{1v}(t)]+[F^N_{2s}(t)-F^N_{2v}(t)], \nonumber
\end{eqnarray}
whereby experimental fact of a creation of true neutral
vector-meson resonances with quantum numbers of the photon in
$e^+e^- \to hadrons$ is in the first approximation taken into
account by a saturation of $F^N_{1s}(t)$, $F^N_{2s}(t)$ with
neutral iso-scalar vector mesons $\omega(782)$, $\phi(1020)$,
$\omega'(1420)$, $\phi'(1680)$, $\omega''(1650)$, $\phi''(2170)$ and
$F^N_{1v}(t)$, $F^N_{2v}(t)$ with neutral iso-vector vector-meson
resonances $\rho(770)$, $\rho'(1450)$, $\rho''(1700)$ in the
corresponding VMD FF parametrization in the zero-width
approximation.

   For the sake of a generality let us consider FF $F(t)$ with a
normalization $F(0)=F_0$, the asymptotic behavior
$F(t)_{t\to\infty}\sim 1/t^m$ and to be saturated with $n$ - true
neutral vector mesons $V_n$. Then in the framework of the standard
$VMD$ model in the zero-width approximation one can write

\begin{equation}
  F(t)=
  \frac{m^2_1}{m^2_1-t}a_1+\frac{m^2_2}{m^2_2-t}a_2+...+\frac{m^2_n}{m^2_n-t}a_n,\label{gvmd}
\end{equation}
where $a_n=(f_{V_nNN}/f_{V_n})$, $f_{V_nNN}$ is the coupling
constant of an interaction of the $n$-th vector-meson with
nucleons and $f_{V_n}$ is the universal vector-meson coupling
constant to be determined numerically from the lepton width of the
$n$-th vector-meson (\ref{lwidth}). Now requiring the
normalization of (\ref{gvmd}) at $t=0$ one gets the following
equation for coupling constant ratios $(f_{iNN}/f_i)$

\begin{equation}
   \sum_{i=1}^n(f_{iNN}/f_i)=F_0.\label{enorm}
\end{equation}

   Then transforming (\ref{gvmd}) into a common denominator one
gets in the numerator polynomial of $t^{n-1}$ degree. In order to
achieve the required asymptotic behavior of $F(t)$ one has to put
some coefficients in the polynomial of the numerator , starting
from the highest degree $t^{n-1}$, to be zero, and one comes to
another $m-1$ equations

\begin{equation}
  \sum_{i=1}^n m^{2r}_i(f_{iNN}/f_i)=0; r=1,2,...,m-1 \label{eeqns}
\end{equation}
for $n$-coupling constant ratios. As a result a solution of the
system of equations (\ref{enorm}),(\ref{eeqns}) will lead $m$
coupling constant ratios to be given through the table masses of
vector mesons and all additional couplig constant ratios
$(f_{m+1}NN/f_{m+1}),...,(f_nNN/f_n)$, which will be considered to
be free parameters of the model.

   The general solution of the system of eqs. (\ref{enorm}),(\ref{eeqns})
for $n>m$ leads \cite{dubnicka} to the FF to be saturated by
$n$-vector-meson resonances in the form suitable for the
unitarization

\begin{eqnarray}
   F(t)&=&F_0\frac{\prod_{j=1}^mm^2_j}{\prod_{j=1}^m(m^2_j-t)}+\label{ffndifm}\\\nonumber
   &+&\sum_{k=m+1}^n\bigg\{\sum_{j=1}^m\frac{m^2_k}{(m^2_k-t)}
   \frac{\prod^m_{j\neq i,j=1}m^2_j}{\prod^m_{j\neq k,j=1}(m^2_j-t)}
   \frac{\prod^m_{j\neq i,j=1}(m^2_j-m^2_k)}{\prod^m_{j\neq
   i,j=1}(m^2_j-m^2_i)}-\\\nonumber
   &-&\frac{\prod_{j=1}^mm^2_j}{\prod_{j=1}^m(m^2_j-t)}\bigg\}(f_{kNN}/f_k)\nonumber
\end{eqnarray}
and for $n=m$

\begin{equation}
   F(t)=F_0\frac{\prod_{j=1}^mm^2_j}{\prod_{j=1}^m(m^2_j-t)}\label{ffneqlm}
\end{equation}
for which the required asymptotic behavior $F(t)_{t\to\infty}\sim
1/t^m$ and for $t=0$ the normalization $F(0)=F_0$ are fulfilled
automatically.

   Now these general results will be applied to the
flavor-independent iso-scalar and iso-vector parts of the Dirac
and Pauli nucleon FFs $F^N_{1s}(t)$, $F^N_{1v}(t)$, $F^N_{2s}(t)$,
$F^N_{2v}(t)$ by means of which all required properties of the
nucleon $U\&A$ electromagnetic structure model like
\begin{itemize}
   \item experimental fact of a creation of unstable vector-meson
resonances in $e^+e^-$ annihilation processes into hadrons

   \item the analytic properties

   \item the reality conditions

   \item the unitarity conditions

   \item normalizations

   \item asymptotic behaviors
 \end{itemize}
are achieved.

   However, before the latter we have to specify normalizations and
asymptotic behaviors of flavor-independent iso-scalar and
iso-vector parts of the Dirac and Pauli nucleon FFs.

   The nucleon EM FFs are normalized as follows
\begin{equation}
   G^p_E(0)=1; G^p_M(0)=\mu_p; G^n_E(0)=0; G^n_M(0)=\mu_n \label{eq5}
\end{equation}
with $\mu_p$ and $\mu_n$ to be proton and neutron magnetic
moments, respectively.

   The asymptotic behaviors of nucleon EM FFs are
\begin{equation}
   G^p_E(t)_{\mid_{t \to \infty}}=G^p_M(t)_{\mid_{t \to \infty}}=G^n_E(t)_{\mid_{t \to \infty}}=G^n_M(t)_{\mid_{t \to
   \infty}}\sim \frac{1}{t^2}.\label{eq6}
\end{equation}

   Then from (\ref{ffrelatios}) one obtains the normalizations of iso-scalar and
iso-vector parts of the Dirac and Pauli nucleon FFs

\begin{equation}
   F^N_{1s}(0)=F^N_{1v}(0)=\frac{1}{2};
   F^N_{2s}(0)=\frac{1}{2}(\mu_p+\mu_n-1);
   F^N_{2v}(0)=\frac{1}{2}(\mu_p-\mu_n-1)\label{eq7}
\end{equation}
and their asymptotic behaviors are

\begin{equation}
   F^N_{1s}(t)_{\mid_{t \to \infty}}=F^N_{1v}(t)_{\mid_{t \to \infty}}\sim\frac{1}{t^2};
   F^N_{2s}(t)_{\mid_{t \to \infty}}=
   F^N_{2v}(t)_{\mid_{t \to \infty}}\sim\frac{1}{t^3}.\label{eq8}
\end{equation}

   If we apply (\ref{ffndifm}) to $F^N_{1s}(t)$ with $m=2$ and
$n=\omega''$, $\phi''$, $\omega'$, $\phi'$, $\omega$, $\phi$ one obtains

\begin{eqnarray}
  F^N_{1s}(t)=\frac{1}{2}\frac{m^2_{\omega''}
  m^2_{\phi''}}{(m^2_{\omega''}-t)(m^2_{\phi''}-t)}+\label{fn1s} \\\nonumber
  +\Bigg\{\frac{m^2_{\phi''} m^2_{\omega'}}{(m^2_{\phi''}-t)(m^2_{\omega'}-t)}
  \frac{(m^2_{\phi''}-m^2_{\omega'})}{(m^2_{\phi''}-m^2_{\omega''})}
  +\frac{m^2_{\omega''} m^2_{\omega'}}{(m^2_{\omega'}-t)(m^2_{\omega''}-t)}
  \frac{(m^2_{\omega''}-m^2_{\omega'})}{(m^2_{\omega''}-m^2_{\phi''})}-\nonumber \\\nonumber
  -\frac{m^2_{\omega''}
  m^2_{\phi''}}{(m^2_{\omega''}-t)(m^2_{\phi''}-t)}\Bigg\}(f^{(1)}_{{\omega'}NN}/f_{\omega'})+ \nonumber \\\nonumber
  +\Bigg\{\frac{m^2_{\phi'} m^2_{\phi''}}{(m^2_{\phi'}-t)(m^2_{\phi''}-t)}
  \frac{(m^2_{\phi''}-m^2_{\phi'})}{(m^2_{\phi''}-m^2_{\omega''})}
  +\frac{m^2_{\omega''} m^2_{\phi'}}{(m^2_{\omega''}-t)(m^2_{\phi'}-t)}
  \frac{(m^2_{\omega''}-m^2_{\phi'})}{(m^2_{\omega''}-m^2_{\phi''})}-\\\nonumber
  -\frac{m^2_{\omega''}
  m^2_{\phi''}}{(m^2_{\omega''}-t)(m^2_{\phi''}-t)}\Bigg\}(f^{(1)}_{{\phi'}NN}/f_{\phi'})+ \nonumber \\\nonumber
  +\Bigg\{\frac{m^2_{\phi''} m^2_{\omega}}{(m^2_{\phi''}-t)(m^2_{\omega}-t)}
  \frac{(m^2_{\phi''}-m^2_{\omega})}{(m^2_{\phi''}-m^2_{\omega''})}
  +\frac{m^2_{\omega} m^2_{\omega''}}{(m^2_{\omega}-t)(m^2_{\omega''}-t)}
  \frac{(m^2_{\omega''}-m^2_{\omega})}{(m^2_{\omega''}-m^2_{\phi''})}-\\\nonumber
  -\frac{m^2_{\omega''}
  m^2_{\phi''}}{(m^2_{\omega''}-t)(m^2_{\phi''}-t)}\Bigg\}(f^{(1)}_{{\omega}NN}/f_{\omega})+ \nonumber \\ \nonumber
  +\Bigg\{\frac{m^2_{\phi} m^2_{\phi''}}{(m^2_{\phi}-t)(m^2_{\phi''}-t)}
  \frac{(m^2_{\phi''}-m^2_{\phi})}{(m^2_{\phi''}-m^2_{\omega''})}
  +\frac{m^2_{\omega''} m^2_{\phi}}{(m^2_{\omega''}-t)(m^2_{\phi}-t)}
  \frac{(m^2_{\omega''}-m^2_{\phi})}{(m^2_{\omega''}-m^2_{\phi''})}-\\\nonumber
  -\frac{m^2_{\omega''}
  m^2_{\phi''}}{(m^2_{\omega''}-t)(m^2_{\phi''}-t)}\Bigg\}(f^{(1)}_{{\phi}NN}/f_{\phi}).\nonumber
\end{eqnarray}

to $F^N_{1v}(t)$ with $m=2$ and $n=\rho''$, $\rho'$, $\rho$ one
obtains

\begin{eqnarray}
  F^N_{1v}(t)=\frac{1}{2}\frac{m^2_{\rho''}
  m^2_{\rho'}}{(m^2_{\rho''}-t)(m^2_{\rho'}-t)}+ \label{fn1v} \\\nonumber
  +\Bigg\{\frac{m^2_{\rho} m^2_{\rho'}}{(m^2_{\rho}-t)(m^2_{\rho'}-t)}
  \frac{(m^2_{\rho'}-m^2_{\rho})}{(m^2_{\rho'}-m^2_{\rho''})}
  +\frac{m^2_{\rho} m^2_{\rho''}}{(m^2_{\rho}-t)(m^2_{\rho''}-t)}
  \frac{(m^2_{\rho''}-m^2_{\rho})}{(m^2_{\rho''}-m^2_{\rho'})}- \nonumber \\\nonumber
  -\frac{m^2_{\rho''}
  m^2_{\rho'}}{(m^2_{\rho''}-t)(m^2_{\rho'}-t)}\Bigg\}(f^{(1)}_{{\rho}NN}/f_{\rho}).\nonumber
\end{eqnarray}

to $F^N_{2s}(t)$ with $m=3$ and $n=\omega'', \phi'', \omega',
\phi', \omega, \phi$ one obtains

\begin{eqnarray}
  F^N_{2s}(t)=\frac{1}{2}(\mu_p+\mu_n-1)\frac{m^2_{\omega''} m^2_{\phi''} m^2_{\omega'}}
  {(m^2_{\omega''}-t)(m^2_{\phi''}-t)(m^2_{\omega'}-t)}+  \label{fn2s}\\ \nonumber
  +\Bigg\{\frac{m^2_{\phi''} m^2_{\phi'}m^2_{\omega'}}{(m^2_{\phi''}-t)(m^2_{\phi'}-t)(m^2_{\omega'}-t)}
  \frac{(m^2_{\phi''}-m^2_{\phi'})(m^2_{\omega'}-m^2_{\phi'})}{(m^2_{\phi''}-m^2_{\omega''})(m^2_{\omega'}-m^2_{\omega''})}+ \\ \nonumber
  +\frac{m^2_{\omega''} m^2_{\omega'} m^2_{\phi'}}{(m^2_{\omega''}-t)(m^2_{\omega'}-t)(m^2_{\phi'}-t)}
  \frac{(m^2_{\omega''}-m^2_{\phi'})(m^2_{\omega'}-m^2_{\phi'})}{(m^2_{\omega''}-m^2_{\phi''})(m^2_{\omega'}-m^2_{\phi''})}+ \\\nonumber
  +\frac{m^2_{\omega''} m^2_{\phi''} m^2_{\phi'}}{(m^2_{\omega''}-t)(m^2_{\phi''}-t)(m^2_{\phi'}-t)}
  \frac{(m^2_{\omega''}-m^2_{\phi'})(m^2_{\phi''}-m^2_{\phi'})}{(m^2_{\omega''}-m^2_{\omega'})(m^2_{\phi''}-m^2_{\omega'})}-\\\nonumber
  -\frac{m^2_{\omega''} m^2_{\phi''} m^2_{\omega'}}{(m^2_{\omega''}-t)(m^2_{\phi''}-t)(m^2_{\omega'}-t)}\Bigg\}
  (f^{(2)}_{{\phi'}NN}/f_{\phi'})+ \\\nonumber
  +\Bigg\{\frac{m^2_{\phi''} m^2_{\omega'}m^2_{\omega}}{(m^2_{\phi''}-t)(m^2_{\omega'}-t)(m^2_{\omega}-t)}
  \frac{(m^2_{\phi''}-m^2_{\omega})(m^2_{\omega'}-m^2_{\omega})}{(m^2_{\phi''}-m^2_{\omega''})(m^2_{\omega'}-m^2_{\omega''})}+ \\\nonumber
  +\frac{m^2_{\omega''} m^2_{\omega'} m^2_{\omega}}{(m^2_{\omega''}-t)(m^2_{\omega'}-t)(m^2_{\omega}-t)}
  \frac{(m^2_{\omega''}-m^2_{\omega})(m^2_{\omega'}-m^2_{\omega})}{(m^2_{\omega''}-m^2_{\phi''})(m^2_{\omega'}-m^2_{\phi''})}+\\\nonumber
  +\frac{m^2_{\omega''} m^2_{\phi''} m^2_{\omega}}{(m^2_{\omega''}-t)(m^2_{\phi''}-t)(m^2_{\omega}-t)}
  \frac{(m^2_{\omega''}-m^2_{\omega})(m^2_{\phi''}-m^2_{\omega})}{(m^2_{\omega''}-m^2_{\omega'})(m^2_{\phi''}-m^2_{\omega'})}-\\ \nonumber
  -\frac{m^2_{\omega''} m^2_{\phi''} m^2_{\omega'}}{(m^2_{\omega''}-t)(m^2_{\phi''}-t)(m^2_{\omega'}-t)}\Bigg\}
  (f^{(2)}_{{\omega}NN}/f_{\omega})+\\\nonumber
  +\Bigg\{\frac{m^2_{\phi''} m^2_{\omega'}m^2_{\phi}}{(m^2_{\phi''}-t)(m^2_{\omega'}-t)(m^2_{\phi}-t)}
  \frac{(m^2_{\phi''}-m^2_{\phi})(m^2_{\omega'}-m^2_{\phi})}{(m^2_{\phi''}-m^2_{\omega''})(m^2_{\omega'}-m^2_{\omega''})}+\\\nonumber
  +\frac{m^2_{\omega''} m^2_{\omega'} m^2_{\phi}}{(m^2_{\omega''}-t)(m^2_{\omega'}-t)(m^2_{\phi}-t)}
  \frac{(m^2_{\omega''}-m^2_{\phi})(m^2_{\omega'}-m^2_{\phi})}{(m^2_{\omega''}-m^2_{\phi''})(m^2_{\omega'}-m^2_{\phi''})}+ \\\nonumber
  +\frac{m^2_{\omega''} m^2_{\phi''} m^2_{\phi}}{(m^2_{\omega''}-t)(m^2_{\phi''}-t)(m^2_{\phi}-t)}
  \frac{(m^2_{\omega''}-m^2_{\phi})(m^2_{\phi''}-m^2_{\phi})}{(m^2_{\omega''}-m^2_{\omega'})(m^2_{\phi''}-m^2_{\omega'})}- \\\nonumber
  -\frac{m^2_{\omega''} m^2_{\phi''} m^2_{\omega'}}{(m^2_{\omega''}-t)(m^2_{\phi''}-t)(m^2_{\omega'}-t)}\Bigg\}
  (f^{(2)}_{{\phi}NN}/f_{\phi}).\nonumber
\end{eqnarray}

   Application of (\ref{ffneqlm}) to $F^N_{2v}(t)$ with $m=3$ and $n=\rho'', \rho',
   \rho$ leads to

\begin{equation}
   F^N_{2v}(t)=\frac{1}{2}(\mu_p -\mu_n-1)\frac{m^2_{\rho''} m^2_{\rho'}
   m^2_{\rho}}{(m^2_{\rho''}-t)(m^2_{\rho'}-t)(m^2_{\rho}-t)}.\label{fn2v}
\end{equation}

  The expressions (\ref{fn1s})-(\ref{fn2v}) are automatically normalized and they govern the asymptotic
behaviors as predicted by the quark model of hadrons.

   An unitarization of the model, i.e. an incorporation of the correct analytic properties
of the nucleon EM FFs, is achieved by the non-linear
transformations
\begin{eqnarray}
t &=& t^s_0 + \frac{4(t^{1s}_{in}-t^s_0)}{[1/V(t)-V(t)]^2};\quad
t = t^v_0 +\frac{4(t^{1v}_{in}-t^v_0)}{[1/W(t)-W(t)]^2};\label{eqtransform}\\
t &=& t^s_0 + \frac{4(t^{2s}_{in}-t^s_0)}{[1/U(t)-U(t)]^2};\quad
t = t^v_0 + \frac{4(t^{2v}_{in}-t^v_0)}{[1/X(t)-X(t)]^2},\nonumber
\end{eqnarray}
respectively and a subsequent inclusion of the nonzero values of
vector-meson widths.

   In non-linear transformations $t^s_0=9m^2_\pi$,
$t^v_0=4m^2_\pi$, $t^{1s}_{in}, t^{1v}_{in}, t^{2s}_{in},
t^{2v}_{in}$ are the square-root branch points, as it is
transparent from the inverse transformations

  \begin{equation}
   V(t)=i\frac{\sqrt{(\frac{t^{1s}_{in}-t^s_0}{t^s_0})^{1/2}+(\frac{t-t^s_0}{t^s_0})^{1/2}}-
   \sqrt{(\frac{t^{1s}_{in}-t^s_0}{t^s_0})^{1/2}-(\frac{t-t^s_0}{t^s_0})^{1/2}}}
   {\sqrt{(\frac{t^{1s}_{in}-t^s_0}{t^s_0})^{1/2}+(\frac{t-t^s_0}{t^s_0})^{1/2}}+
   \sqrt{(\frac{t^{1s}_{in}-t^s_0}{t^s_0})^{1/2}-(\frac{t-t^s_0}{t^s_0})^{1/2}}}\label{eq9}
\end{equation}
and similarly for $W(t)$, $U(t)$, and $X(t)$, which map the
corresponding four-sheeted Riemann surfaces always into one $V-,
W-, U-, X-$ plane.

  Practically let us demonstrate the unitarization on the case of
the Dirac iso-scalar FF (\ref{fn1s}).
   The non-linear transformation $t = t^s_0 + \frac{4(t^{1s}_{in}-t^s_0)}{[1/V(t)-V(t)]^2}$
implies also the relations $m^2_r
=t^s_0+\frac{4(t^{1s}_{in}-t^s_0)}{[1/V_{r0}-V_{r0}]^2}$ and $0
=t^s_0 + \frac{4(t^{1s}_{in}-t^s_0)}{[1/V_N-V_N]^2}.$

   Then every term
\begin{equation}
  \frac{m^2_r}{m^2_r-t}\equiv
  \frac{m^2_r-0}{m^2_r-t}=\bigg(\frac{1-V^2}{1-V^2_N}\bigg)^2
  \frac{(V_N-V_{r0})(V_N+V_{r0})(V_N-1/V_{r0})(V_N+1/V_{r0})}
  {(V-V_{r0})(V+V_{r0})(V-1/V_{r0})(V+1/V_{r0})}\label{eq10}
\end{equation}
in (\ref{fn1s}) is factorized into asymptotic term and on the so-called
finite-energy term (for $\mid t \mid \to \infty$ it turns out to
be a real constant) giving a resonant behavior around $t=m^2_r$.

   One can prove
\begin{itemize}
  \item[1.] \quad if\quad $m^2_r-\Gamma^2_r/4<t_{in} \Rightarrow
  V_{r0}=-V^*_{r0}$

\item[2.]  \quad if\quad $m^2_r-\Gamma^2_r/4>t_{in} \Rightarrow
  V_{r0}=1/V^*_{r0}$
\end{itemize}
which lead in the case 1. to the expression

\begin{equation}
  \frac{m^2_r}{m^2_r-t}=\bigg(\frac{1-V^2}{1-V^2_N}\bigg)^2
  \frac{(V_N-V_{r0})(V_N-V^*_{r0})(V_N-1/V_{r0})(V_N-1/V^*_{r0})}
  {(V-V_{r0})(V-V^*_{r0})(V-1/V_{r0})(V-1/V^*_{r0})}\label{eq11}
\end{equation}
and in the case 2. to the following expression

\begin{equation}
  \frac{m^2_r}{m^2_r-t}=\bigg(\frac{1-V^2}{1-V^2_N}\bigg)^2
  \frac{(V_N-V_{r0})(V_N-V^*_{r0})(V_N+V_{r0})(V_N+V^*_{r0})}
  {(V-V_{r0})(V-V^*_{r0})(V+V_{r0})(V+V^*_{r0})}.\label{eq12}
\end{equation}

   Lastly, introducing the non-zero width of the resonance by a
substitution

\begin{equation}
   m^2_r \rightarrow (m_r-i\Gamma_r/2)^2\label{eq13}
\end{equation}
i.e. simply one has to rid of "0" in sub-indices of the previous
two expressions, one gets:

in the 1. case

\begin{eqnarray}
  \frac{m^2_r}{m^2_r-t} &\rightarrow& \bigg(\frac{1-V^2}{1-V^2_N}\bigg)^2
  \frac{(V_N-V_{r})(V_N-V^*_{r})(V_N-1/V_{r})(V_N-1/V^*_{r})}
  {(V-V_{r})(V-V^*_{r})(V-1/V_{r})(V-1/V^*_{r})}=\\\nonumber
  &=& \bigg(\frac{1-V^2}{1-V^2_N}\bigg)^2 L_r(V)\label{eq14}
\end{eqnarray}

and in the 2. case

\begin{eqnarray}
  \frac{m^2_r}{m^2_r-t} &\rightarrow& \bigg(\frac{1-V^2}{1-V^2_N}\bigg)^2
  \frac{(V_N-V_{r})(V_N-V^*_{r})(V_N+V_{r})(V_N+V^*_{r})}
  {(V-V_{r})(V-V^*_{r})(V+V_{r})(V+V^*_{r})}=\\\nonumber
  &=& \bigg(\frac{1-V^2}{1-V^2_N}\bigg)^2 H_r(V).\label{eq15}
\end{eqnarray}

   Then for every iso-scalar and iso-vector Dirac and Pauli
FF one obtains just one analytic and smooth from $-\infty$ to
$+\infty$ function in the forms

\begin{eqnarray}
  F^N_{1s}[V(t)]=\Bigg(\frac{1-V^2}{1-V^2_N}\Bigg)^4\Bigg\{\frac{1}{2}H_{\omega''}(V)H_{\phi''}(V)+\\\nonumber
  +\Bigg[H_{\phi''}(V)H_{\omega'}(V)\frac{(C^{1s}_{\phi''}-C^{1s}_{\omega'})}{(C^{1s}_{\phi''}-C^{1s}_{\omega''})}+
  H_{\omega''}(V)H_{\omega'}(V)\frac{(C^{1s}_{\omega''}-C^{1s}_{\omega'})}{(C^{1s}_{\omega''}-C^{1s}_{\phi''})}-\\\nonumber
  -H_{\omega''}(V)H_{\phi''}(V)\Bigg](f^{(1)}_{\omega'NN}/f_{\omega'})+\\\nonumber
  +\Bigg[H_{\phi''}(V)H_{\phi'}(V)\frac{(C^{1s}_{\phi''}-C^{1s}_{\phi'})}{(C^{1s}_{\phi''}-C^{1s}_{\omega''})}+
  H_{\omega''}(V)H_{\phi'}(V)\frac{(C^{1s}_{\omega''}-C^{1s}_{\phi'})}{(C^{1s}_{\omega''}-C^{1s}_{\phi''})}-\\\nonumber
  -H_{\omega''}(V)H_{\phi''}(V)\Bigg](f^{(1)}_{\phi'NN}/f_{\phi'})+\\\nonumber
  +\Bigg[H_{\phi''}(V)L_{\omega}(V)\frac{(C^{1s}_{\phi''}-C^{1s}_{\omega})}{(C^{1s}_{\phi''}-C^{1s}_{\omega''})}+
  H_{\omega''}(V)L_{\omega}(V)\frac{(C^{1s}_{\omega''}-C^{1s}_{\omega})}{(C^{1s}_{\omega''}-C^{1s}_{\phi''})}-\\\nonumber
  -H_{\omega''}(V)H_{\phi''}(V)\Bigg](f^{(1)}_{\omega NN}/f_{\omega})+\\\nonumber
  +\Bigg[H_{\phi''}(V)L_{\phi}(V)\frac{(C^{1s}_{\phi''}-C^{1s}_{\phi})}{(C^{1s}_{\phi''}-C^{1s}_{\omega''})}+
  H_{\omega''}(V)L_{\phi}(V)\frac{(C^{1s}_{\omega''}-C^{1s}_{\phi})}{(C^{1s}_{\omega''}-C^{1s}_{\phi''})}-\\\nonumber
  -H_{\omega''}(V)H_{\phi''}(V)\Bigg](f^{(1)}_{\phi NN}/f_{\phi})\Bigg\}\label{eq16}
\end{eqnarray}
dependent on 5 free physically interpretable parameters,

$(f^{(1)}_{\omega'NN}/f_{\omega'})$, $(f^{(1)}_{\phi'NN}/f_{\phi'})$,
$(f^{(1)}_{\omega NN}/f_{\omega})$, $(f^{(1)}_{\phi NN}/f_{\phi})$,
$t^{1s}_{in}$

\begin{eqnarray}
  F^N_{1v}[W(t)]=\Bigg(\frac{1-W^2}{1-W^2_N}\Bigg)^4\Bigg\{\frac{1}{2}L_\rho(W)L_{\rho'}(W)+\\\nonumber
  +\Bigg[L_{\rho'}(W)L_{\rho''}(W)\frac{(C^{1v}_{\rho'}-C^{1v}_{\rho''})}{(C^{1v}_{\rho'}-C^{1v}_\rho)}+
  L_\rho(W)L_{\rho''}(W)\frac{(C^{1v}_\rho-C^{1v}_{\rho''})}{(C^{1v}_\rho-C^{1v}_{\rho'})}-\\\nonumber
  -L_\rho(W)L_{\rho'}(W)\Bigg](f^{(1)}_{\rho NN}/f_{\rho})\Bigg\}\label{eq17}
\end{eqnarray}
dependent on 2 free physically interpretable parameters
$(f^{(1)}_{\rho NN}/f_{\rho})$ and $t^{1v}_{in}$

\begin{eqnarray}
  F^N_{2s}[U(t)]=\Bigg(\frac{1-U^2}{1-U^2_N}\Bigg)^6\Bigg\{\frac{1}{2}(\mu_p+\mu_n-1)H_{\omega''}(U)H_{\phi''}(U)H_{\omega'}(U)+\\\nonumber
  +\Bigg[H_{\phi''}(U)H_{\omega'}(U)H_{\phi'}(U)\frac{(C^{2s}_{\phi''}-C^{2s}_{\phi'})(C^{2s}_{\omega'}-C^{2s}_{\phi'})}
  {(C^{2s}_{\phi''}-C^{2s}_{\omega''})(C^{2s}_{\omega'}-C^{2s}_{\omega''})}+\\\nonumber
  +H_{\omega''}(U)H_{\omega'}(U)H_{\phi'}(U)\frac{(C^{2s}_{\omega''}-C^{2s}_{\phi'})(C^{2s}_{\omega'}-C^{2s}_{\phi'})}
  {(C^{2s}_{\omega''}-C^{2s}_{\phi''})(C^{2s}_{\omega'}-C^{2s}_{\phi''})}+\\\nonumber
  +H_{\omega''}(U)H_{\phi''}(U)H_{\phi'}(U)\frac{(C^{2s}_{\omega''}-C^{2s}_{\phi'})(C^{2s}_{\phi''}-C^{2s}_{\phi'})}
  {(C^{2s}_{\omega''}-C^{2s}_{\omega'})(C^{2s}_{\phi''}-C^{2s}_{\omega'})}-\\\nonumber
  -H_{\omega''}(U)H_{\phi''}(U)H_{\omega'}(U)\Bigg](f^{(2)}_{\phi'NN}/f_{\phi'})+\\\nonumber
  +\Bigg[H_{\phi''}(U)H_{\omega'}(U)L_{\omega}(U)\frac{(C^{2s}_{\phi''}-C^{2s}_{\omega})(C^{2s}_{\omega'}-C^{2s}_{\omega})}
  {(C^{2s}_{\phi''}-C^{2s}_{\omega''})(C^{2s}_{\omega'}-C^{2s}_{\omega''})}+\\\nonumber
  +H_{\omega''}(U)H_{\omega'}(U)L_{\omega}(U)\frac{(C^{2s}_{\omega''}-C^{2s}_{\omega})(C^{2s}_{\omega'}-C^{2s}_{\omega})}
  {(C^{2s}_{\omega''}-C^{2s}_{\phi''})(C^{2s}_{\omega'}-C^{2s}_{\phi''})}+\\\nonumber
  +H_{\omega''}(U)H_{\phi''}(U)L_{\omega}(U)\frac{(C^{2s}_{\omega''}-C^{2s}_{\omega})(C^{2s}_{\phi'}-C^{2s}_{\omega})}
  {(C^{2s}_{\omega''}-C^{2s}_{\omega'})(C^{2s}_{\phi''}-C^{2s}_{\omega'})}-\\\nonumber
  -H_{\omega''}(U)H_{\phi''}(U)H_{\omega'}(U)\Bigg](f^{(2)}_{\omega NN}/f_{\omega})+\\\nonumber
  +\Bigg[H_{\phi''}(U)H_{\omega'}(U)L_{\phi}(U)\frac{(C^{2s}_{\phi''}-C^{2s}_{\phi})(C^{2s}_{\omega'}-C^{2s}_{\phi})}
  {(C^{2s}_{\phi''}-C^{2s}_{\omega''})(C^{2s}_{\omega'}-C^{2s}_{\omega''})}+\\\nonumber
  +H_{\omega''}(U)H_{\omega'}(U)L_{\phi}(U)\frac{(C^{2s}_{\omega''}-C^{2s}_{\phi})(C^{2s}_{\omega'}-C^{2s}_{\phi})}
  {(C^{2s}_{\omega''}-C^{2s}_{\phi''})(C^{2s}_{\omega'}-C^{2s}_{\phi''})}+\\\nonumber
  +H_{\omega''}(U)H_{\phi''}(U)L_{\phi}(U)\frac{(C^{2s}_{\omega''}-C^{2s}_{\phi})(C^{2s}_{\phi''}-C^{2s}_{\phi})}
  {(C^{2s}_{\omega''}-C^{2s}_{\omega'})(C^{2s}_{\phi''}-C^{2s}_{\omega'})}-\\\nonumber
  -H_{\omega''}(U)H_{\phi''}(U)H_{\omega'}(U)\Bigg](f^{(2)}_{\phi NN}/f_{\phi})\Bigg\}\label{eq18}
\end{eqnarray}
dependent on 4 free physically interpretable parameters

$(f^{(2)}_{\phi'NN}/f_{\phi'})$, $(f^{(2)}_{\omega NN}/f_{\omega})$,
$(f^{(2)}_{\phi NN}/f_{\phi}), t^{2s}_{in}$

\begin{eqnarray}
  F^N_{2v}[X(t)]=\Bigg(\frac{1-X^2}{1-X^2_N}\Bigg)^6\Bigg\{\frac{1}{2}(\mu_p-\mu_n-1)L_\rho(U)L_{\rho'}(U)H_{\rho''}(U)\Bigg\}\label{eq19}
\end{eqnarray}
dependent on 1 free physically interpretable parameter
$t^{2v}_{in}$, where

\begin{eqnarray}
  L_r(V)=\frac{(V_N-V_r)(V_N-V^*_r)(V_N-1/V_r)(V_N-1/V^*_r)}{(V-V_r)(V-V^*_r)(V-1/V_r)(V-1/V^*_r)},\\\label{eq19}
  C^{1s}_r=\frac{(V_N-V_r)(V_N-V^*_r)(V_N-1/V_r)(V_N-1/V^*_r)}{-(V_r-1/V_r)(V_r-1/V^*_r)}, r=\omega, \phi \nonumber
\end{eqnarray}

\begin{eqnarray}
  H_l(V)=\frac{(V_N-V_l)(V_N-V^*_l)(V_N+V_l)(V_N+V^*_l)}{(V-V_l)(V-V^*_l)(V+V_l)(V+V^*_l)},\\\label{eq20}
  C^{1s}_l=\frac{(V_N-V_l)(V_N-V^*_l)(V_N+V_l)(V_N+V^*_l)}{-(V_l-1/V_l)(V_l-1/V^*_l)}, l=
  \omega'', \phi'', \omega', \phi' \nonumber
\end{eqnarray}

\begin{eqnarray}
  L_k(W)=\frac{(W_N-W_k)(W_N-W^*_k)(W_N-1/W_k)(W_N-1/W^*_k)}{(W-W_k)(W-W^*_k)(W-1/W_k)(W-1/W^*_k)},\\ \label{eq21}
  C^{1v}_k=\frac{(W_N-W_k)(W_N-W^*_k)(W_N-1/W_k)(W_N-1/W^*_k)}{-(W_k-1/W_k)(W_k-1/W^*_k)}, k=\rho'',
  \rho', \rho \nonumber
\end{eqnarray}

\begin{eqnarray}
  L_r(U)=\frac{(U_N-U_r)(U_N-U^*_r)(U_N-1/U_r)(U_N-1/U^*_r)}{(U-U_r)(U-U^*_r)(U-1/U_r)(U-1/U^*_r)},\\ \label{eq22}
  C^{2s}_r=\frac{(U_N-U_r)(U_N-U^*_r)(U_N-1/U_r)(U_N-1/U^*_r)}{-(U_r-1/U_r)(U_r-1/U^*_r)}, r=\omega, \phi \nonumber
\end{eqnarray}

\begin{eqnarray}
  H_l(U)=\frac{(U_N-U_l)(U_N-U^*_l)(U_N+U_l)(U_N+U^*_l)}{(U-U_l)(U-U^*_l)(U+U_l)(U+U^*_l)},\\ \label{eq23}
  C^{2s}_l=\frac{(U_N-U_l)(U_N-U^*_l)(U_N+U_l)(U_N+U^*_l)}{-(U_l-1/U_l)(U_l-1/U^*_l)}, l=
  \omega'', \phi'', \omega', \phi' \nonumber
\end{eqnarray}

\begin{eqnarray}
  L_k(X)=\frac{(X_N-X_k)(X_N-X^*_k)(X_N-1/X_k)(X_N-1/X^*_k)}{(X-X_k)(X-X^*_k)(X-1/X_k)(X-1/X^*_k)},\\\label{eq24}
  C^{2v}_k=\frac{(X_N-X_k)(X_N-X^*_k)(X_N-1/X_k)(X_N-1/X^*_k)}{-(X_k-1/X_k)(X_k-1/X^*_k)}, k=\rho', \rho \nonumber
\end{eqnarray}

\begin{eqnarray}
  H_{\rho''}(X)=\frac{(X_N-X_{\rho''})(X_N-X^*_{\rho''})(X_N+X_{\rho''})(X_N+X^*_{\rho''})}
  {(X-X_{\rho''})(X-X^*_{\rho''})(X+X_{\rho''})(X+X^*_{\rho''})},\\ \label{eq25}
  C^{2v}_{\rho''}=\frac{(X_N-X_{\rho''})(X_N-X^*_{\rho''})(X_N+X_{\rho''})(W_X+X^*_{\rho''})}
  {-(X_{\rho''}-1/X_{\rho''})(X_{\rho''}-1/X^*_{\rho''})}.\nonumber
\end{eqnarray}
since in a fitting procedure of all existing data by means of this
version of the $U\&A$ nucleon EM structure model simultaneously
one finds

\begin{eqnarray}
  (m^2_\omega-\Gamma^2_\omega/4)<t^{1s}_{in};(m^2_\phi-\Gamma^2_\phi/4)<t^{1s}_{in};\\\nonumber
  (m^2_{\omega'}-\Gamma^2_{\omega'}/4)>t^{1s}_{in};(m^2_{\phi'}-\Gamma^2_{\phi'}/4)>t^{1s}_{in};\\\nonumber
  (m^2_{\omega''}-\Gamma^2_{\omega''}/4)>t^{1s}_{in};(m^2_{\phi''}-\Gamma^2_{\phi''}/4)>t^{1s}_{in};\nonumber
\end{eqnarray}

\begin{eqnarray}
  (m^2_\rho-\Gamma^2_\rho/4)<t^{1v}_{in};(m^2_{\rho'}-\Gamma^2_{\rho'}/4)<t^{1v}_{in};
  (m^2_{\rho''}-\Gamma^2_{\rho''}/4)<t^{1v}_{in};\label{eq26}
\end{eqnarray}

\begin{eqnarray}
  (m^2_\omega-\Gamma^2_\omega/4)<t^{2s}_{in};(m^2_\phi-\Gamma^2_\phi/4)<t^{2s}_{in};\\\nonumber
  (m^2_{\omega'}-\Gamma^2_{\omega'}/4)>t^{2s}_{in};(m^2_{\phi'}-\Gamma^2_{\phi'}/4)>t^{2s}_{in};\\\nonumber
  (m^2_{\omega''}-\Gamma^2_{\omega''}/4)>t^{2s}_{in};(m^2_{\phi''}-\Gamma^2_{\phi''}/4)>t^{2s}_{in};\\\nonumber
\end{eqnarray}

\begin{eqnarray}
  (m^2_\rho-\Gamma^2_\rho/4)<t^{2v}_{in};(m^2_{\rho'}-\Gamma^2_{\rho'}/4)<t^{2v}_{in};
  (m^2_{\rho''}-\Gamma^2_{\rho''}/4)>t^{2v}_{in}.\label{eq27}
\end{eqnarray}

\section{Results of the analysis of existing nucleon EM FF data}

   We have collected 534 reliable experimental points on the nucleon
EM structure from more than 40 independent experiments as they are
represented graphically in Figs.~\ref{fig1}-\ref{fig11}. They have been analysed
simultaneously by means of the 9 resonance $U\&A$ model of the
nucleon EM structure as formulated in the previous Section. The
minimum of the $\chi^2$=2 214 has been achieved with the values of
12 free parameters of the model with a clear physical meaning as
they are presented in {\bf Table 1.},

\newpage

{\bf Table 1.} The numerical values of the free parameters of the
nucleon $U\&A$ EM structure model, respecting SU(3) symmetry,
formulated in the previous section

\begin{eqnarray}\nonumber
    t^{1s}_{in}&=& (1.0442 \pm 0.0200) GeV^2\\\nonumber
    t^{2s}_{in}&=& (1.0460 \pm 0.1399) GeV^2\\\nonumber
    t^{1v}_{in}&=& (2.9506 \pm 0.5326) GeV^2\\\nonumber
    t^{2v}_{in}&=& (2.3449 \pm 0.7656) GeV^2\\\nonumber
    (f^{(1)}_{\omega NN}/f_{\omega}) &=& (1.5717 \pm 0.0022)\\\nonumber
    (f^{(1)}_{\phi NN}/f_{\phi}) &=& (-1.1247 \pm 0.0011)\\\nonumber
    (f^{(1)}_{\omega' NN}/f_{\omega'}) &=& (0.0418 \pm 0.0065)\\\nonumber
    (f^{(1)}_{\phi' NN}/f_{\phi'}) &=& (0.1879 \pm 0.0010)\\\nonumber
    (f^{(2)}_{\omega NN}/f_{\omega}) &=& (-0.2096 \pm 0.0067)\\\nonumber
    (f^{(2)}_{\phi NN}/f_{\phi}) &=& (0.2657 \pm 0.0067)\\\nonumber
    (f^{(2)}_{\phi' NN}/f_{\phi'}) &=& (0.1781 \pm 0.0029)\\\nonumber
    (f^{(1)}_{\rho NN}/f_{\rho}) &=& (0.3747 \pm
    0.0022)\\\nonumber\label{fitvalues}
\end{eqnarray}
whereby the results are not very sensitive on the position of the
effective inelastic thresholds $t^{1s}_{in}$, $t^{2s}_{in}$,
$t^{1v}_{in}$, $t^{2v}_{in}$.

\begin{figure}
\includegraphics[scale=0.6]{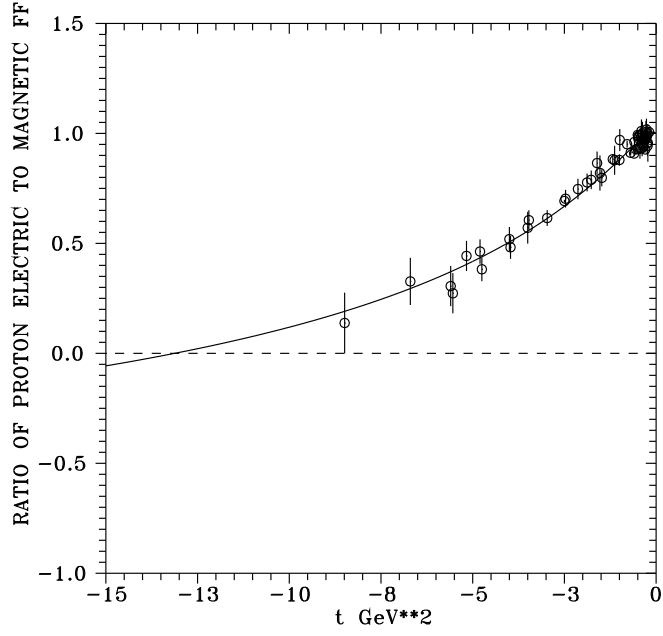}
\caption{Prediction of proton electric to magnetic FFs ratio
behavior in space-like region by $U\&A$ model respecting $SU(3)$
symmetry and its comparison with existing data}.
\label{fig12}
\end{figure}

\begin{figure}
\includegraphics[scale=0.6]{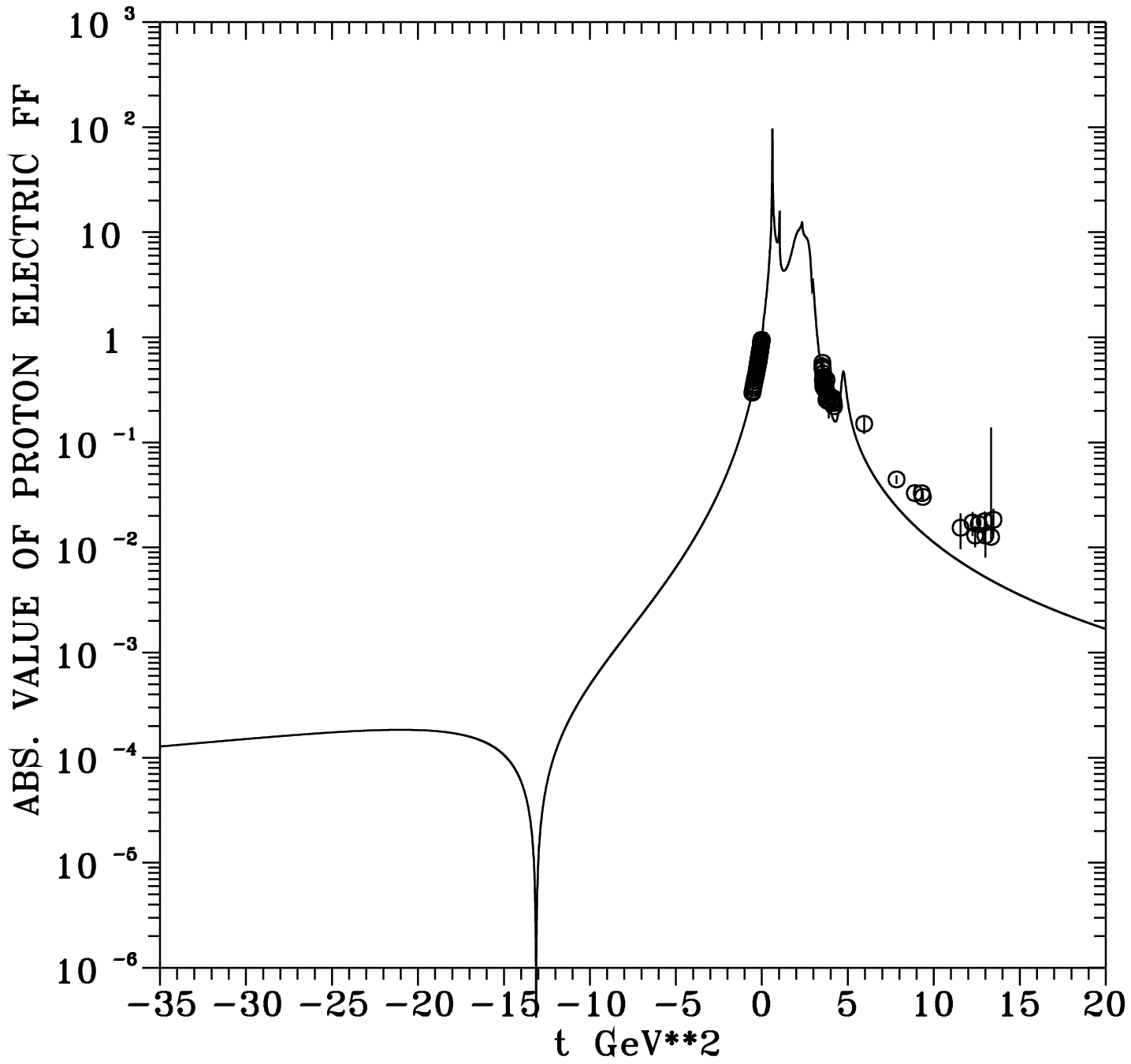}
\caption{Prediction of proton electric FF behavior by $U\&A$ model
respecting $SU(3)$ symmetry and its comparison with existing
data.}
\label{fig13}
\end{figure}

\begin{figure}
\includegraphics[scale=0.6]{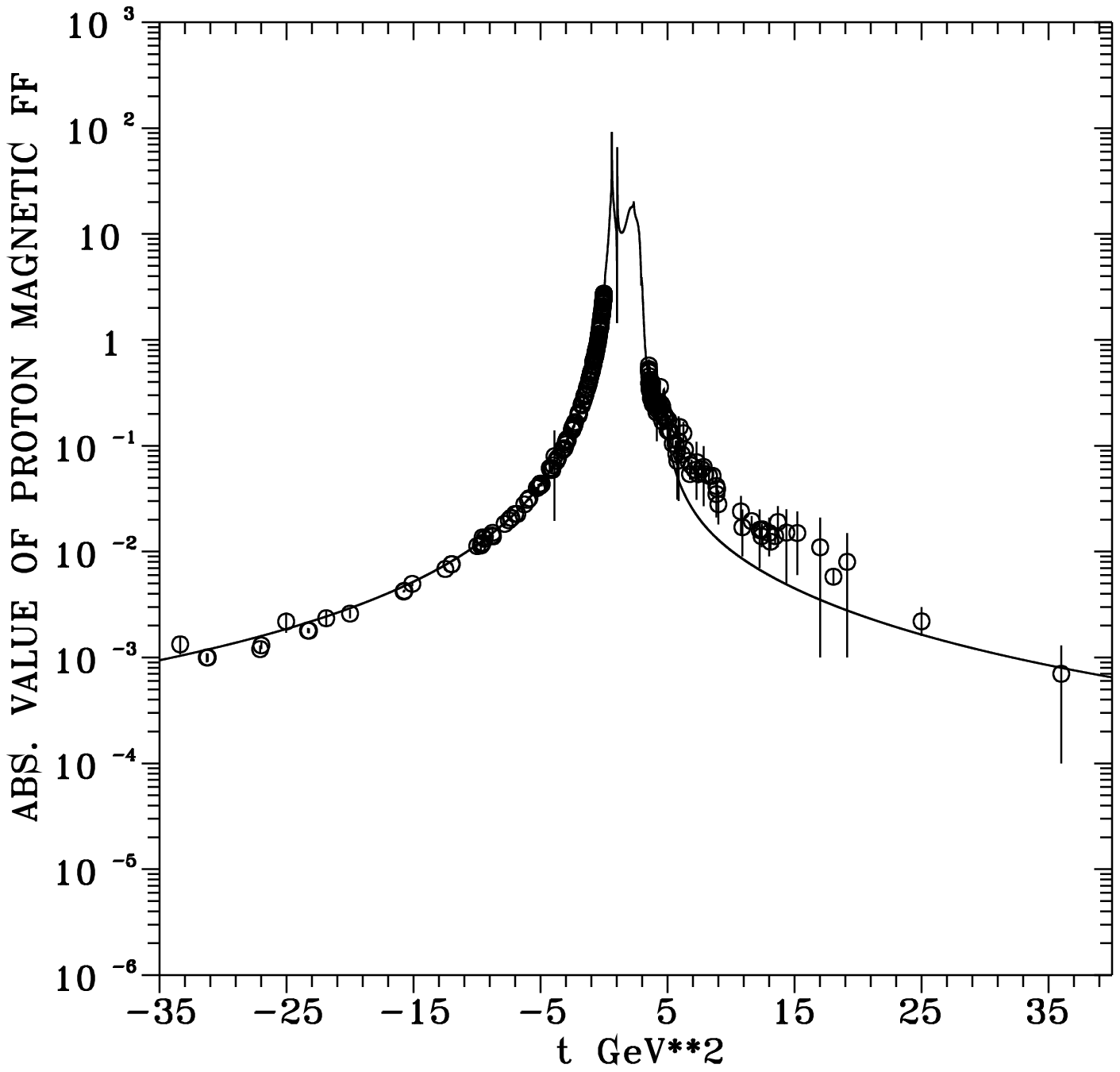}
\caption{Prediction of proton magnetic FF behavior by $U\&A$ model
respecting $SU(3)$ symmetry and its comparison with existing
data.}
\label{fig14}
\end{figure}

\begin{figure}
\includegraphics[scale=0.6]{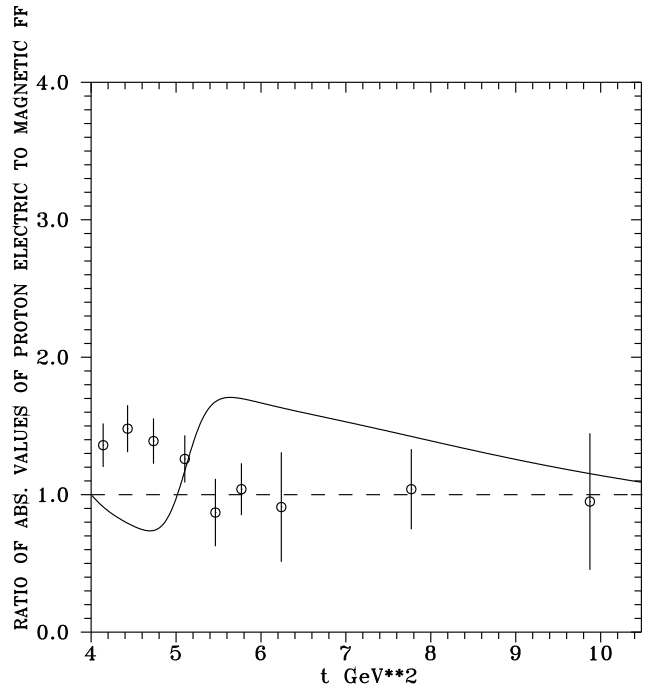}
\caption{Prediction of the absolute value of proton electric to
magnetic FFs ratio behavior in time-like region by $U\&A$ model
respecting $SU(3)$ symmetry and its comparison with existing
data.}
\label{fig15}
\end{figure}

\begin{figure}
\includegraphics[scale=0.6]{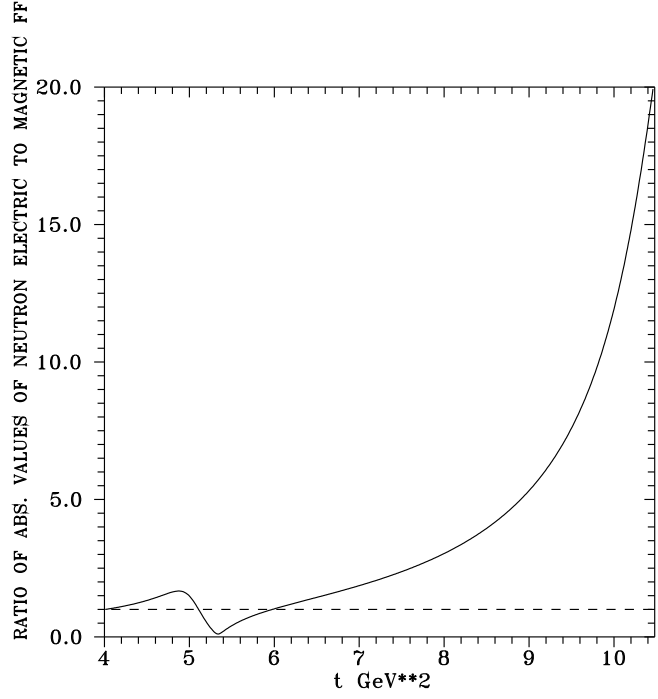}
\caption{Prediction of the absolute value of neutron electric to
magnetic FFs ratio behavior in time-like region by $U\&A$ model
respecting $SU(3)$ symmetry.}
\label{fig16}
\end{figure}

\begin{figure}
\includegraphics[scale=0.6]{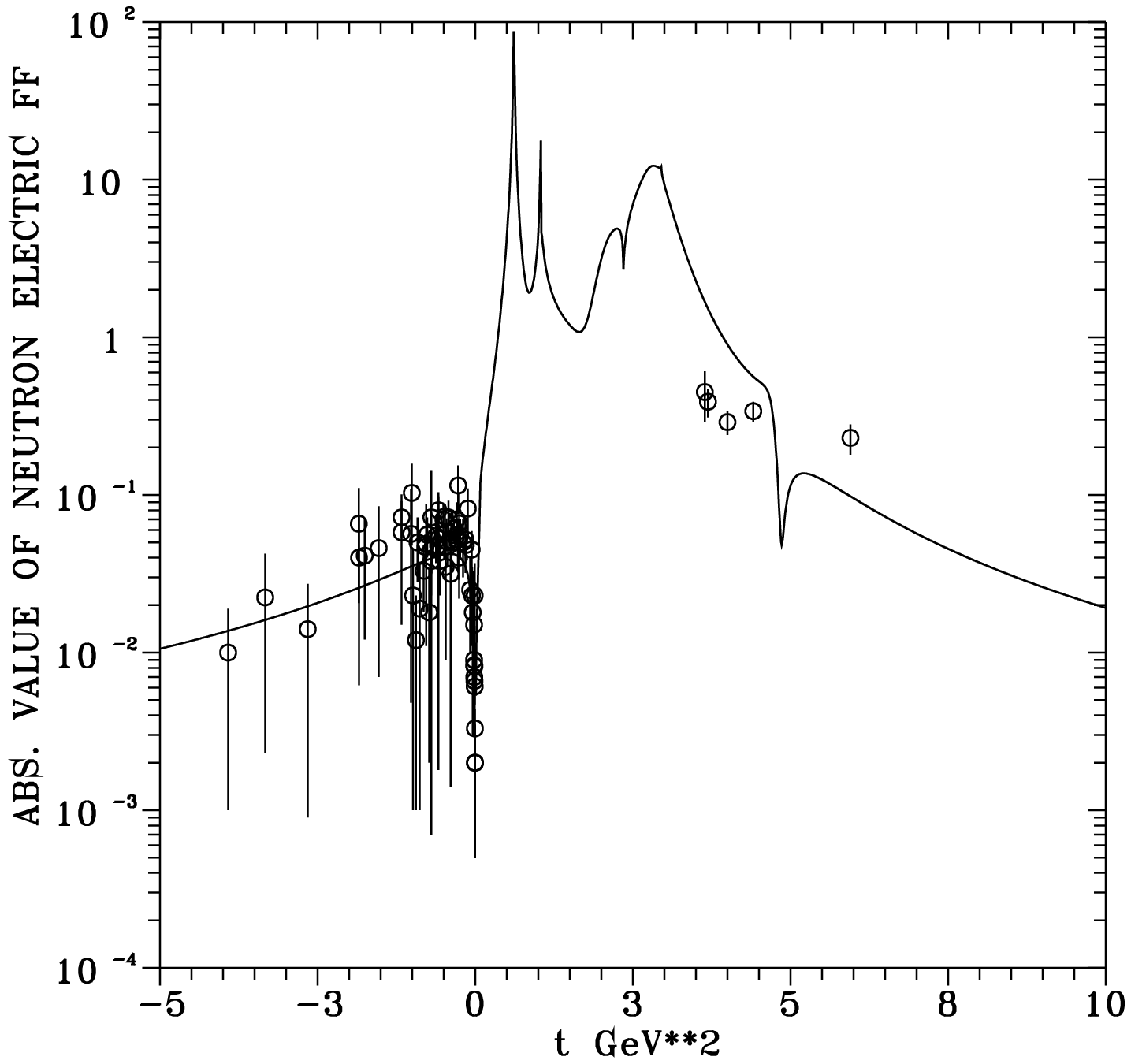}
\caption{Prediction of neutron electric FF behavior by $U\&A$
model respecting $SU(3)$ symmetry and its comparison with existing
the data.}
\label{fig17}
\end{figure}

\begin{figure}
\includegraphics[scale=0.6]{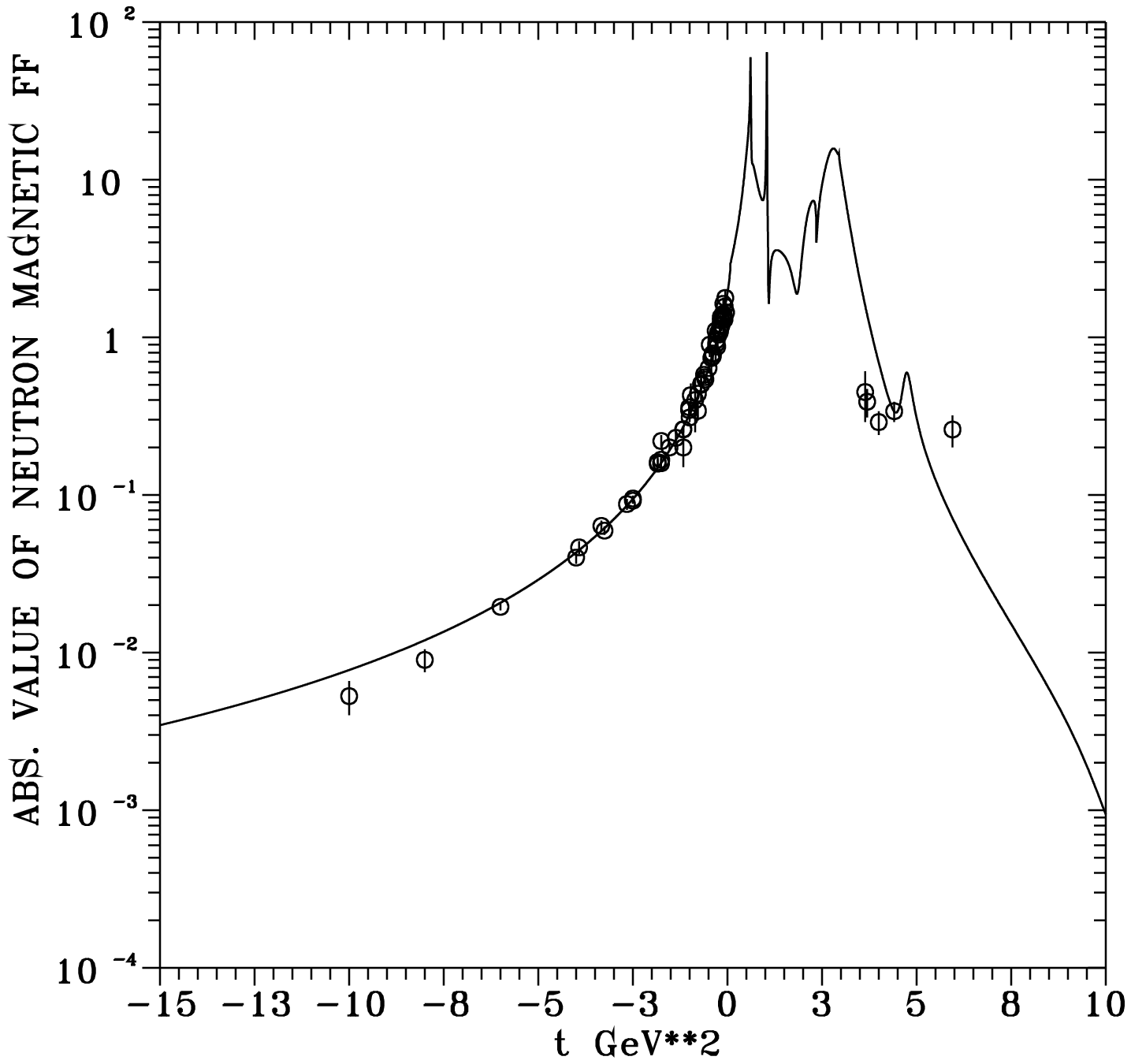}
\caption{Prediction of neutron magnetic FF behavior by $U\&A$
model respecting $SU(3)$ symmetry and its comparison with
existing data.}
\label{fig18}
\end{figure}

\begin{figure}
\includegraphics[scale=0.6]{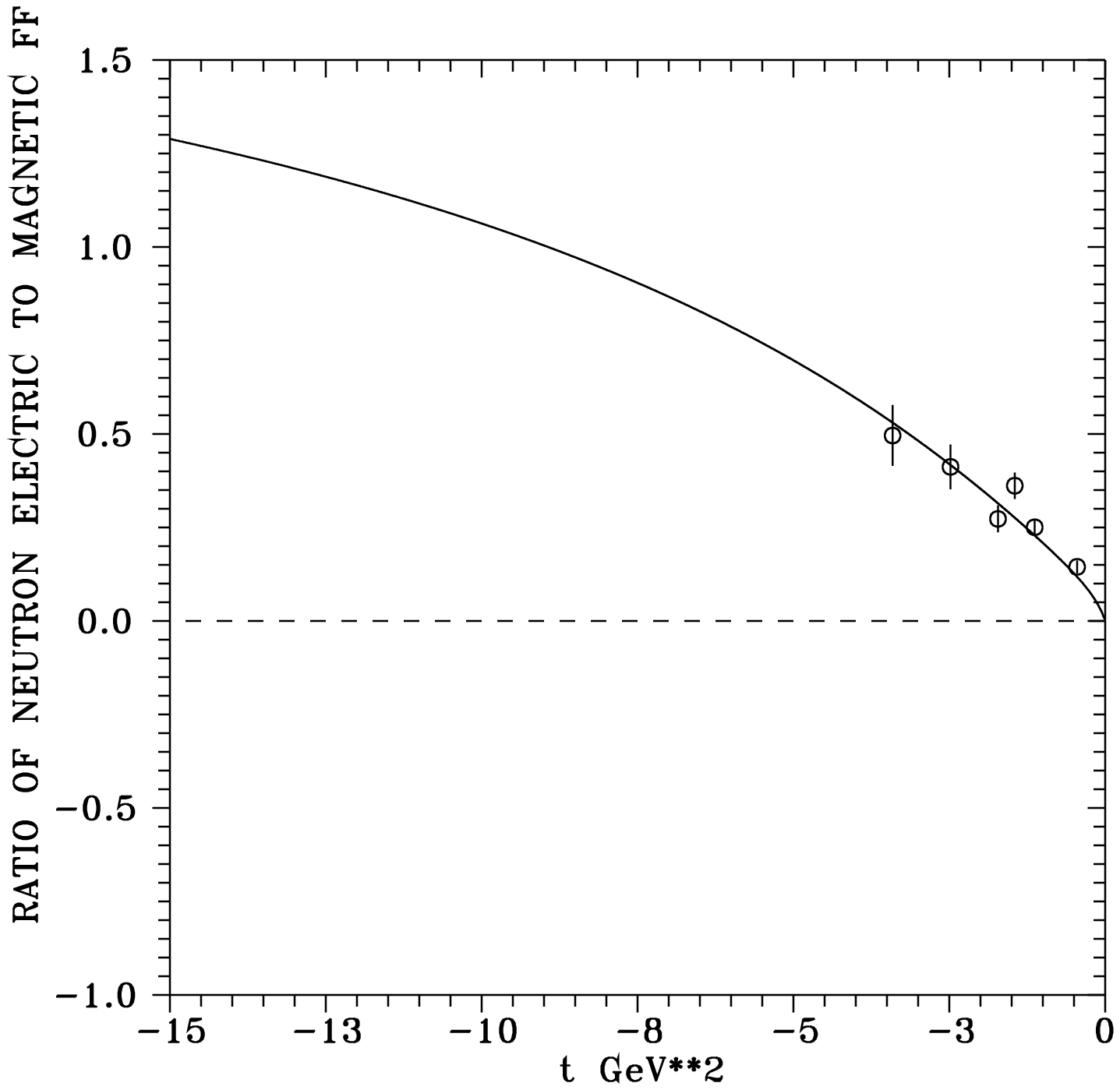}
\caption{Prediction of neutron electric to magnetic FFs ratio
behavior in space-like region by $U\&A$ model respecting $SU(3)$
symmetry and its comparison with existing data.}
\label{fig19}
\end{figure}

   The corresponding description of the data by means of this
$U\&A$ model with numerical values of parameters given in Table 1.
is graphically shown in Figs.~\ref{fig12}-\ref{fig19}.

   Of course one could not expect that a description of such
gigantic set of data to be obtained from so much independent
experiments, every of them to be charged with corresponding
statistical and systematical errors, will be consistent with rules
of a standard statistics. We have been able to reduce the total
$\chi^2$ on 522 degrees of freedom only to the value 4.24.

   The results of the analysis can be summarized as follows:
\begin{itemize}
       \item a perfect description (see Fig.~\ref{fig12}) of the most reliable nucleon EM
       structure data, i.e. the data on the ratio
       $\mu_pG^p_E(t)/G^p_M(t)$ in the space-like region to be
       obtained in polarization experiments, is achieved.

       \item description of all other existing data (see Figs.~\ref{fig13}-\ref{fig19}) is quite
       reasonable too, besides an inconsistency of the data on neutron
       EM FFs in the time-like region with all other data on nucleon EM FFs,
       indicating that the total-cross section of $e^+e^- \to n \bar
       n$ is considerably larger than it has been found in FENICE experiment
       \cite{antonelli} in Frascati. So, we are coming to the same
       conclusions as it was pointed out by one of the the authors in papers
       \cite{dubnicka2}, \cite{dubnicka3} published more than 25 years ago.

       \item again the existence of the zero of the proton electric FF $G^p_E(t)$
       approximately at $t_z=-13$ GeV$^2$ is confirmed, which has been predicted
       in the paper \cite{adamuscin} for the first time.

       \item electric and magnetic mean square charge radii of the nucleons are
       determined to be $\quad$ $<r^2_{Ep}>=(0.7182 \pm 0.0369)$ fm$^2$,$\quad $
       $<r^2_{Mp}>=(0.7573 \pm 0.0133)$ fm$^2$,$\quad$ $\quad $ $<r^2_{En}>=(-0.1162 \pm 0.0369)$ fm$^2$,
       $<r^2_{Mn}>=(0.8312 \pm 0.0195)$ fm$^2$, whereby the
       obtained electric root mean square charge radius of the
       proton is consistent with the value $<r_{Ep}>=0.84184\pm 0.00067$ fm determined in
       the muon hydrogen atom spectroscopy \cite{pohl} and in this way the
       electric proton charge radius puzzle is definitely solved. Recently
       to the same conclusions came also Ulf.-G.Meissner with
       collaborators in \cite{meissner}.

       \item electric mean square charge radius of the neutron is found to be almost identical with
       the value given by Rev. of Part. Physics \cite{olive}.
\end{itemize}

\begin{figure}
\includegraphics[scale=0.6]{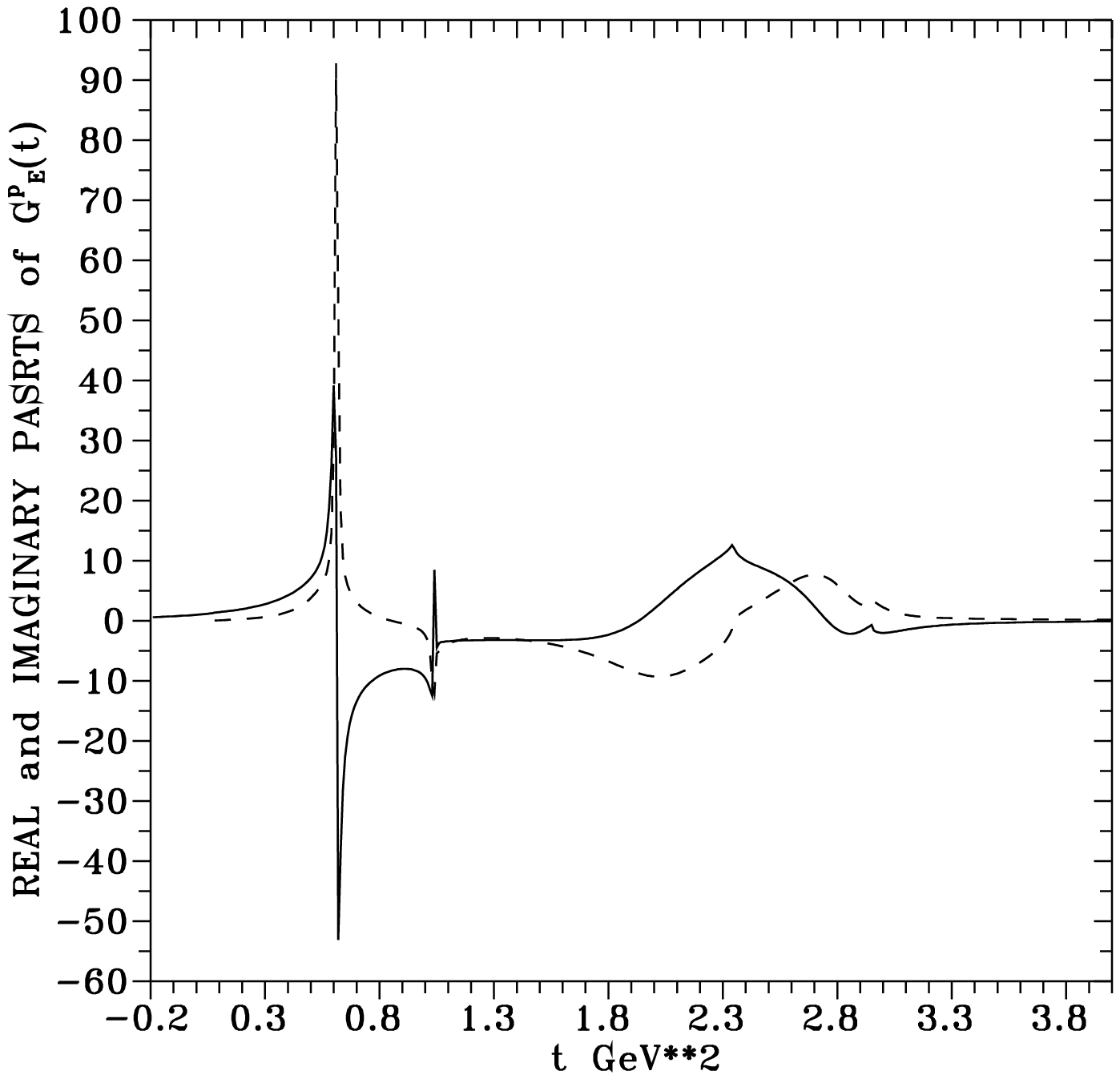}
\caption{Prediction of real (solid line) and imaginary (dashed
line) parts of the proton electric FF by $U\&A$ model respecting
$SU(3)$ symmetry.}
\label{fig20}
\end{figure}

\begin{figure}
\includegraphics[scale=0.6]{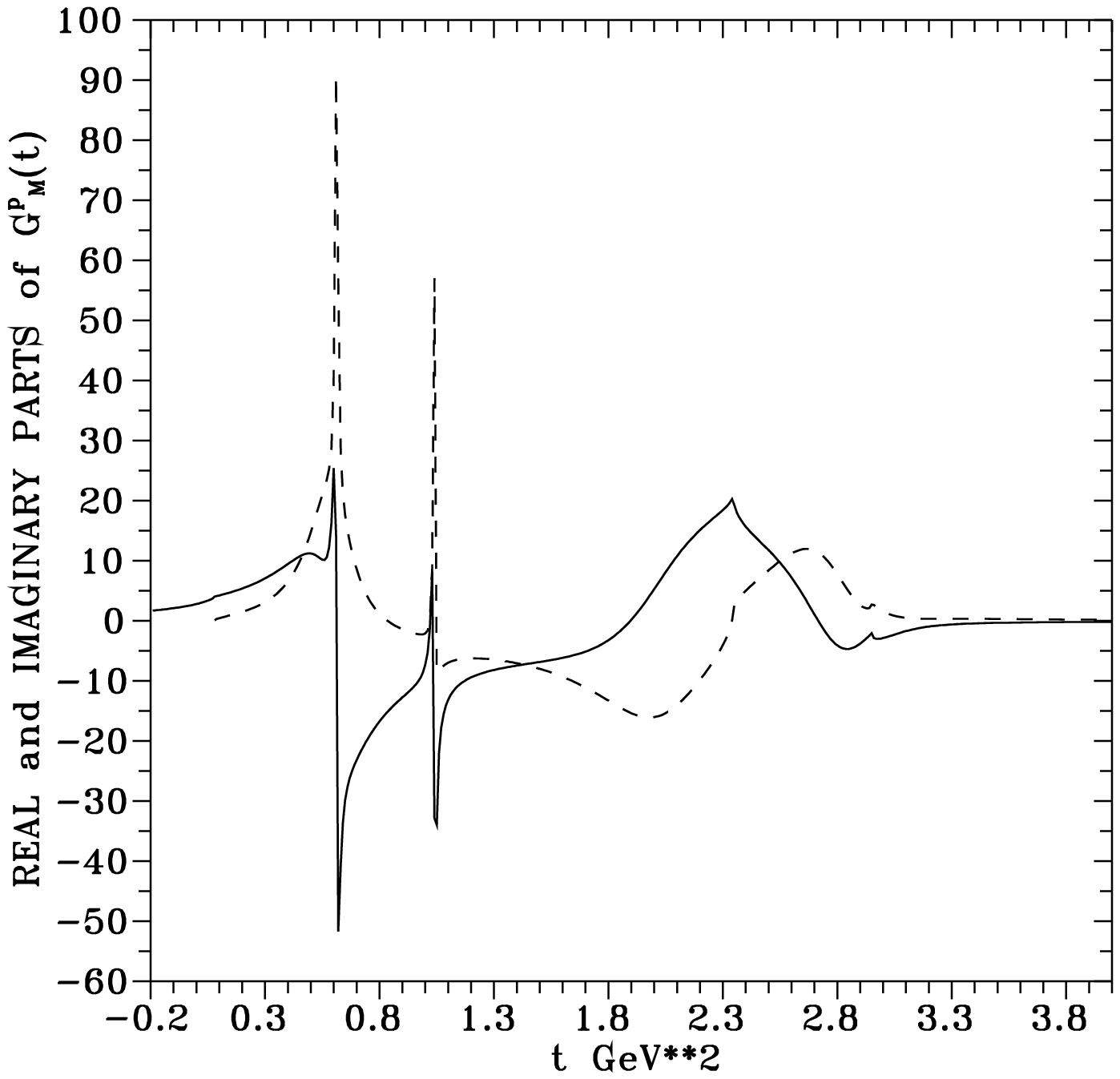}
\caption{Prediction of real (solid line) and imaginary (dashed
line) parts of the proton magnetic FF by $U\&A$ model respecting
$SU(3)$ symmetry.}
\label{fig21}
\end{figure}

\begin{figure}
\includegraphics[scale=0.6]{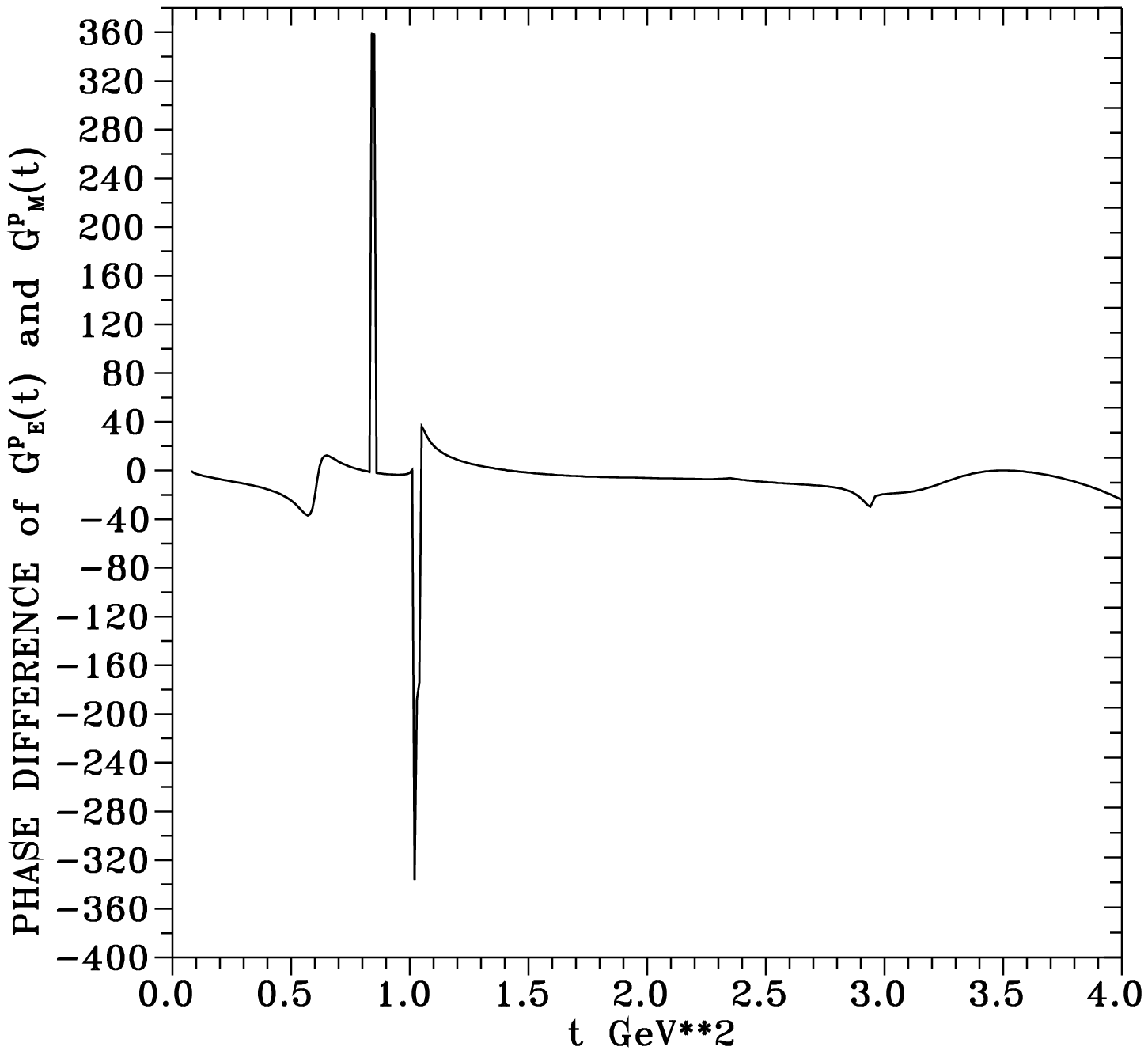}
\caption{Prediction of a phase difference of the proton electric
and magnetic FFs by $U\&A$ model respecting $SU(3)$ symmetry.}
\label{fig22}
\end{figure}

\begin{figure}
\includegraphics[scale=0.6]{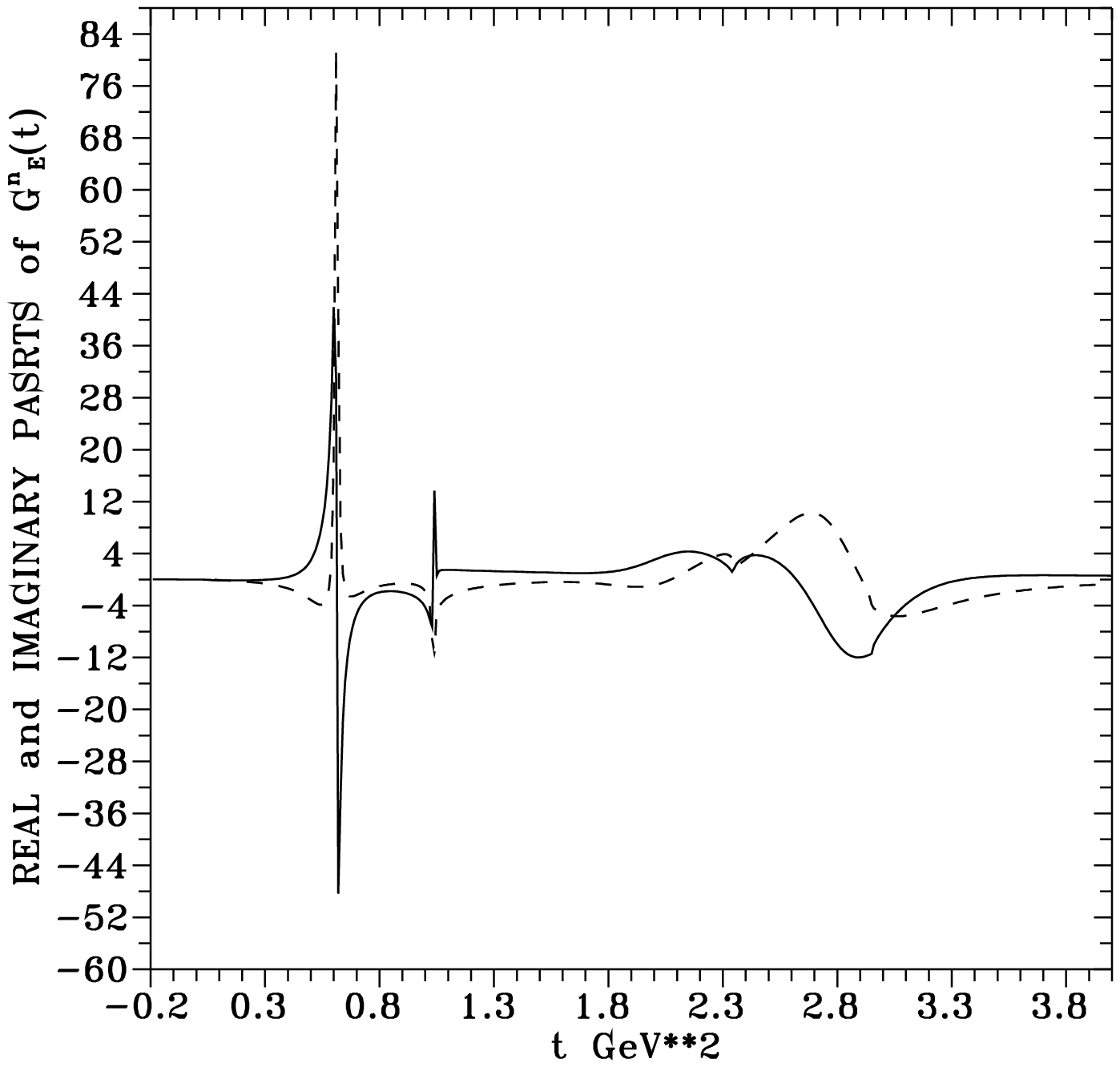}
\caption{Prediction of real (solid line) and imaginary (dashed
line) parts of the neutron electric FF by $U\&A$ model respecting
$SU(3)$ symmetry.}
\label{fig23}
\end{figure}

\begin{figure}
\includegraphics[scale=0.6]{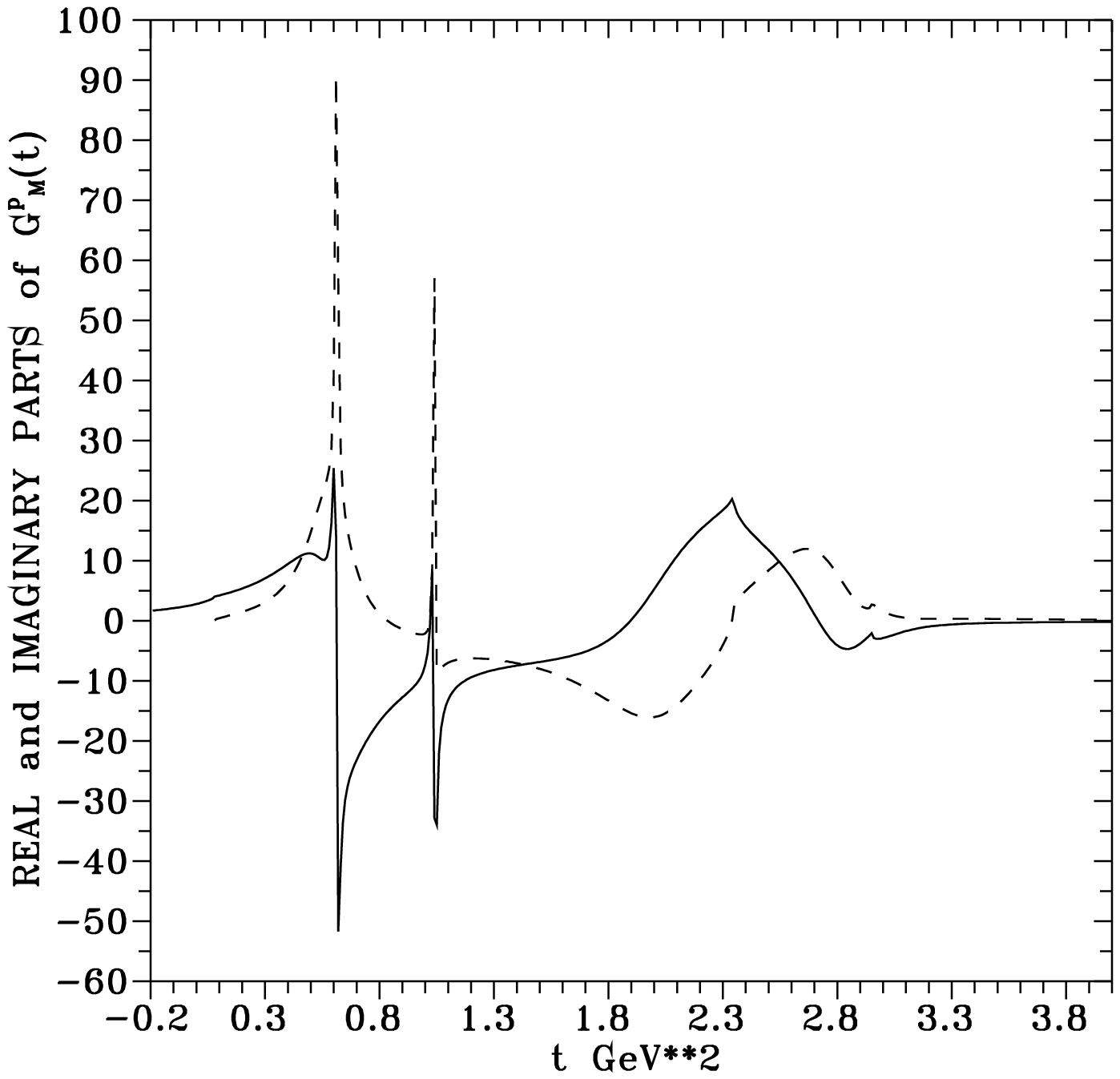}
\caption{Prediction of real (solid line) and imaginary (dashed
line) parts of the neutron magnetic FF by $U\&A$ model respecting
$SU(3)$ symmetry.}
\label{fig24}
\end{figure}

\begin{figure}
\includegraphics[scale=0.6]{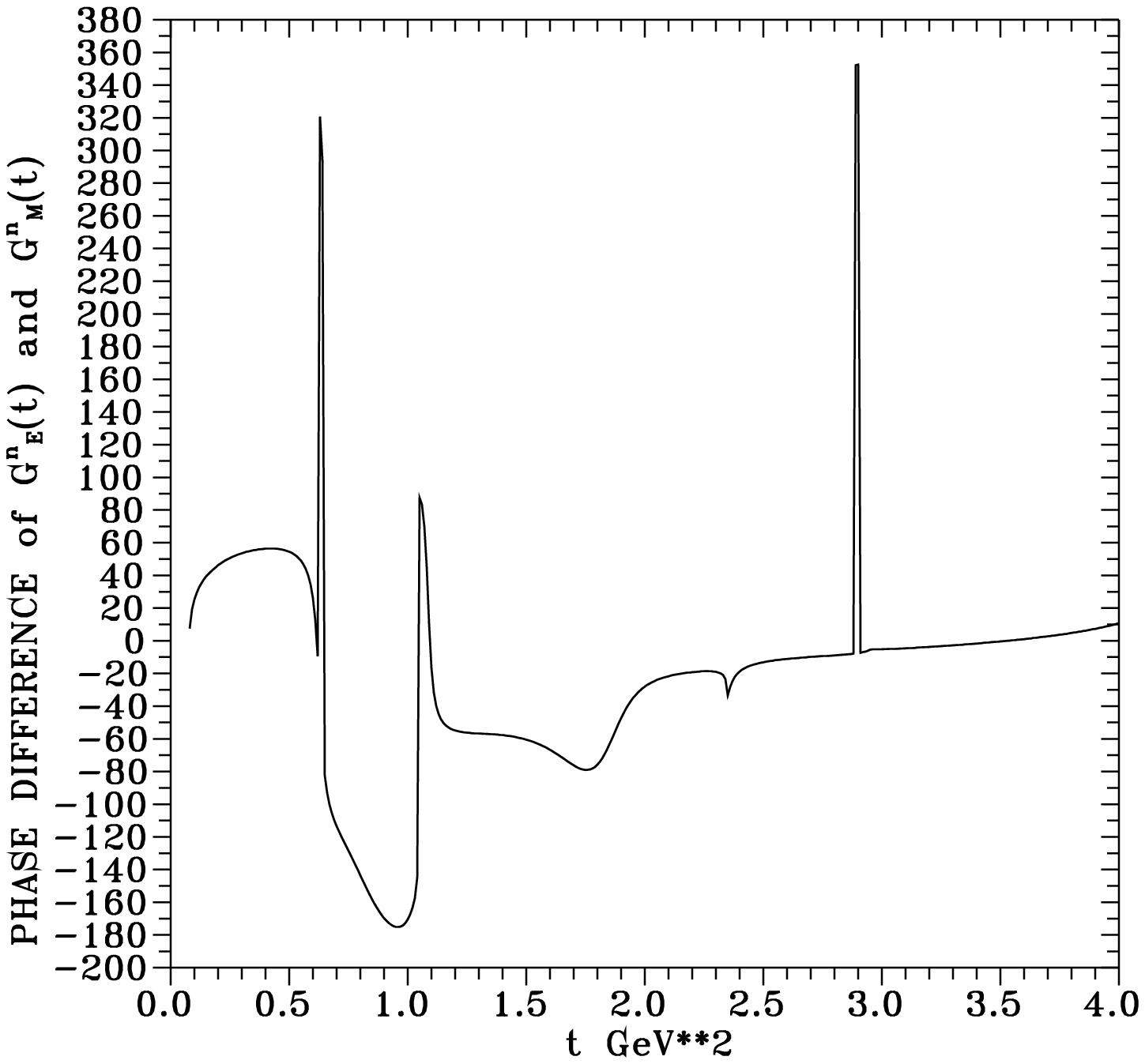}
\caption{Prediction of a phase difference of the neutron electric
and magnetic FFs by $U\&A$ model respecting $SU(3)$ symmetry.}
\label{fig25}
\end{figure}

   In Figs.~\ref{fig20}-\ref{fig25} it is clearly demonstrated that the nucleon EM FFs
represented by the $U\&A$ model indeed fulfil the reality
condition $G^*(t)=G(t^*)$, i.e. they all are real functions from
$-\infty$ up to the lowest branch point $t_0=4 m^2_\pi$ on the
positive real axis.

   The imaginary parts of all nucleon EM FFs are different from zero just
only from the lowest branch point at $t=0.0784 GeV^2$ to $+\infty$
and their behaviors are given by the unitarity conditions of the
corresponding FFs.

\section{Numerical values of the $f^F$, $f^D$, $f^S$ and  $f^{F'}$, $f^{D'}$, $f^{S'}$ coupling constants}

   The $SU(3)$ invariant Lagrangian (2) of vector-meson nonet
 $\rho^-$, $\rho^0$, $\rho^+$, $K^{*-}$, $K^{*0}$, $\bar K^{*0}$, $K^{*+}$, $\omega$, $\phi$
with $1/2^+$ octet baryons $p$, $n$, $\Lambda$, $\Sigma^+$, $\Sigma^0$,
$\Sigma^-$, $\Xi^0$, $\Xi^-$ provides the following expressions

\begin{eqnarray}
f^{(1)}_{\rho NN}&=&\frac{1}{2}(f^D_{1}+f^F_{1})\label{eqfron}\\\nonumber
f^{(1)}_{\omega NN}&=&\frac{1}{\sqrt{2}}f^S_{1}\cos\theta  -
\frac{1}{2\sqrt{3}}(3f^F_{1}-f^D_{1})\sin\theta\\\nonumber
f^{(1)}_{\phi NN}&=&\frac{1}{\sqrt{2}}f^S_{1}\sin\theta  +
\frac{1}{2\sqrt{3}}(3f^F_{1}-f^D_{1})\cos\theta\\\nonumber
\end{eqnarray}

and

\begin{eqnarray}
f^{(2)}_{\rho NN}&=&\frac{1}{2}(f^D_{2}+f^F_{2})\label{eqfonn}\\\nonumber
f^{(2)}_{\omega NN}&=&\frac{1}{\sqrt{2}}f^S_{2}\cos\theta  -
\frac{1}{2\sqrt{3}}(3f^F_{2}-f^D_{2})\sin\theta\\\nonumber
f^{(2)}_{\phi NN}&=&\frac{1}{\sqrt{2}}f^S_{2}\sin\theta  +
\frac{1}{2\sqrt{3}}(3f^F_{2}-f^D_{2})\cos\theta\\\nonumber
\end{eqnarray}
where angle $\theta=43.8^0$ and it is given by the Gell-Mann-Okubo quadratic
mass formula
\begin{eqnarray}
m^2_{\phi(1020)} \cos^2\theta + m^2_{\omega(782)} \sin^2\theta =
\frac{4m^2_{K^*(892)}-m^2_{\rho(770)}}{3}.\label{eqf1}
\end{eqnarray}

From the $SU(3)$ invariant Lagrangian of the first excited vector-meson nonet
 $\rho^{-'}$, $\rho^{0'}$, $\rho^{+'}$, $K^{*-'}$, $K^{*0'}$, $\bar K^{*0'}$, $K^{*+'}$, $\omega'$, $\phi'$
with $1/2^+$ octet baryons $p$, $n$, $\Lambda$, $\Sigma^+$, $\Sigma^0$,
$\Sigma^-$, $\Xi^0$, $\Xi^-$

\begin{eqnarray}
L_{V'B\bar{B}}=\frac{i}{\sqrt{2}}f^F[\bar{B}^\alpha_\beta
\gamma_\mu B^\beta_\gamma- \bar{B}^\beta_\gamma\gamma_\mu
B^\alpha_\beta](V'_\mu)^\gamma_\alpha+\label{lagrangianexc}\\\nonumber
+\frac{i}{\sqrt{2}}f^D[\bar{B}^\beta_\gamma \gamma_\mu
B^\alpha_\beta+
\bar{B}^\alpha_\gamma\gamma_\mu B^\beta_\gamma](V'_\mu)^\gamma_\alpha+\\ \nonumber
+\frac{i}{\sqrt{2}}f^S \bar{B}^\alpha_\beta \gamma_\mu
B^\beta_\alpha\omega^{0'}_\mu \nonumber
\end{eqnarray}
another two systems of expressions

\begin{eqnarray}
f^{(1)}_{\rho' NN}&=&\frac{1}{2}(f^{D'}_{1}+f^{F'}_{1})\label{eq1ffn}\\\nonumber
f^{(1)}_{\omega' NN}&=&\frac{1}{\sqrt{2}}f^{S'}_{1}\cos\theta'  -
\frac{1}{2\sqrt{3}}(3f^{F'}_{1}-f^{D'}_{1})\sin\theta'\\\nonumber
f^{(1)}_{\phi' NN}&=&\frac{1}{\sqrt{2}}f^{S'}_{1}\sin\theta'  +
\frac{1}{2\sqrt{3}}(3f^{F'}_{1}-f^{D'}_{1})\cos\theta'\nonumber
\end{eqnarray}

and

\begin{eqnarray}
f^{(2)}_{\rho' NN}&=&\frac{1}{2}(f^{D'}_{2}+f^{F'}_{2})\label{eq2ffn}\\\nonumber
f^{(2)}_{\omega' NN}&=&\frac{1}{\sqrt{2}}f^{S'}_{2}\cos\theta'  -
\frac{1}{2\sqrt{3}}(3f^{F'}_{2}-f^{D'}_{2})\sin\theta'\\\nonumber
f^{(2)}_{\phi' NN}&=&\frac{1}{\sqrt{2}}f^{S'}_{2}\sin\theta'  +
\frac{1}{2\sqrt{3}}(3f^{F'}_{2}-f^{D'}_{2})\cos\theta'\\\nonumber
\end{eqnarray}
are obtained, where angle $\theta'=50.3^0$ and it is again given by the
Gell-Mann-Okubo quadratic mass formula
\begin{eqnarray}
m^2_{\phi'(1680)} \cos^2\theta' + m^2_{\omega'(1420)} \sin^2\theta' =
\frac{4m^2_{K^{*'}(1680)}-m^2_{\rho'(1450)}}{3},\label{eqf2}
\end{eqnarray}
however, with the first excited states of the corresponding
particles

   So, if one knows numerical values of the coupling constants on the
left-hand side of the expressions (\ref{eqfron}),(\ref{eqfonn}) and (\ref{eq1ffn}),(\ref{eq2ffn}) one can
find numerical values of all $f^F$, $f^D$, $f^S$ and $f^{F'}$, $f^{D'}$,
$f^{S'}$ coupling constants in both $SU(3)$ invariant Lagrangians
(\ref{lagrangian}) and (\ref{lagrangianexc}), needed for a prediction of EM FFs behaviors of all
$1/2$ octet hyperons $\Lambda$, $\Sigma^+$, $\Sigma^0$, $\Sigma^-$,
$\Xi^0$, $\Xi^-$.

   For (\ref{eqfron}) and (\ref{eqfonn}) it is straightforward to determine them from
the numerical values in Table 1., besides $(f^{(2)}_{\rho
NN}/f_\rho)$ which is calculated by means of the relation

\begin{eqnarray}
 (f^{(2)}_{\rho NN}/f_\rho)=\frac{1}{2}(\mu_p-\mu_n-1)\frac{C^{2v}_{\rho''}C^{2v}_{\rho'}}
 {(C^{2v}_{\rho'}-C^{2v}_{\rho})(C^{2v}_{\rho''}-C^{2v}_{\rho})}=2.8956,\label{eqf4}
\end{eqnarray}
and the values $f_\rho=4.9582$, $f_\omega=17.0620$, $f_\phi=-13.4428$
to be calculated from existing data \cite{olive} on lepton width
$\Gamma(V \to e^+e^-)$ by means of the relation $\Gamma(V \to
e^+e^-)=\frac{\alpha^2 m_V}{3}{(\frac{f^2_V}{4\pi})}^{-1}$.

   The results are
\begin{eqnarray}
f^{(1)}_{\rho NN}&=&1.8578\label{eqf59} \\\nonumber f^{(1)}_{\omega
NN}&=&26.8163\\\nonumber f^{(1)}_{\phi NN}&=&15.1191\nonumber
\end{eqnarray}

and

\begin{eqnarray}
f^{(2)}_{\rho NN}&=&14.3570\label{eqf60}\\\nonumber f^{(2)}_{\omega
NN}&=&-3.5762\\\nonumber f^{(2)}_{\phi NN}&=&-3.5718\nonumber
\end{eqnarray}

  However, for excited vector meson coupling constants in (\ref{eq1ffn}) and (\ref{eq2ffn}) a big
problem appeared. There are no data (see \cite{olive}) on
$\Gamma(V \to e^+e^-)$ for $\omega'(1420), \rho'(1450),
\phi'(1680)$ in order to determine $f_{\rho'}$, $f_{\omega'}$ and
$f_{\phi'}$. The first two are calculated from lepton widths
estimated by Donnachie and Clegg \cite{donnachieclegg}
$f_{\rho'}=13.6491$, $f_{\omega'}=47.6022$ and the third one for
$\phi'$ is determined to be $f_{\phi'}=-33.6598$ from the
relations
$f^2_{\rho'}:f^2_{\omega'}:f^2_{\phi'}=\frac{1}{9}:1:\frac{1}{2}$
following (see \cite{bramon}) from the quark structure of the
corresponding vector mesons and the electric charges of the three
constituent quarks from which these vector mesons are compound.

   Having numerical values of $f_{\rho'}, f_{\omega'}, f_{\phi'}$ and
calculating missing in Table 1. coupling constant ratios for
excited vector mesons by the relations

\begin{eqnarray}
 (f^{(1)}_{\rho'NN}/f_{\rho'})=\frac{1}{2}\frac{C^{1v}_{\rho''}}{(C^{1v}_{\rho''}-
 C^{1v}_{\rho'})}-\frac{(C^{1v}_{\rho''}-
 C^{1v}_{\rho})}{(C^{1v}_{\rho''}-
 C^{1v}_{\rho'})}(f^{(1)}_{\rho NN}/f_{\rho})=0.7635\label{eqf61}
\end{eqnarray}

\begin{eqnarray}
 (f^{(2)}_{\rho' NN}/f_{\rho'})=-\frac{1}{2}(\mu_p-\mu_n-1)\frac{C^{2v}_{\rho''}C^{2v}_{\rho}}
 {(C^{2v}_{\rho''}-C^{2v}_{\rho'})(C^{2v}_{\rho'}-C^{2v}_{\rho})}=-1.3086,\label{eqf62}
\end{eqnarray}

\begin{eqnarray}
 (f^{(2)}_{\omega' NN}/f_{\omega'})=\frac{1}{2}(\mu_p+\mu_n-1)\frac{C^{2s}_{\omega''}C^{2s}_{\phi''}}
 {(C^{2s}_{\phi''}-C^{2s}_{\omega'})(C^{2s}_{\omega''}-C^{2s}_{\omega'})}-\label{eqf63}\\\nonumber
 -\frac{(C^{2s}_{\phi''}-C^{2s}_{\omega})(C^{2s}_{\omega''}-C^{2s}_{\omega})}
 {(C^{2s}_{\phi''}-C^{2s}_{\omega'})(C^{2s}_{\omega''}-C^{2s}_{\omega'})}(f^{(2)}_{\omega
 NN}/f_{\omega})-\\\nonumber
 -\frac{(C^{2s}_{\phi''}-C^{2s}_{\phi})(C^{2s}_{\omega''}-C^{2s}_{\phi})}
 {(C^{2s}_{\phi''}-C^{2s}_{\omega'})(C^{2s}_{\omega''}-C^{2s}_{\omega'})}(f^{(2)}_{\phi
 NN}/f_{\phi})-\\\nonumber
 -\frac{(C^{2s}_{\phi''}-C^{2s}_{\phi'})(C^{2s}_{\omega''}-C^{2s}_{\phi'})}
 {(C^{2s}_{\phi''}-C^{2s}_{\omega'})(C^{2s}_{\omega''}-C^{2s}_{\omega'})}(f^{(2)}_{\phi'
 NN}/f_{\phi'})=-0.5771\nonumber
\end{eqnarray}
one obtains the results
\begin{eqnarray}
f^{(1)}_{\rho' NN}&=&10.4211\label{eqf64}\\\nonumber f^{(1)}_{\omega'
NN}&=&1.9900\\\nonumber f^{(1)}_{\phi' NN}&=&-6.3247\\\nonumber
\end{eqnarray}

and

\begin{eqnarray}
f^{(2)}_{\rho' NN}&=&-17.8612\label{eqf65}\\\nonumber f^{(2)}_{\omega'
NN}&=&-27.4712\\\nonumber f^{(2)}_{\phi' NN}&=&-5.9948\\\nonumber
\end{eqnarray}

Now, by a solution of the systems of algebraic equations (\ref{eqfron}),
(\ref{eqfonn}), (\ref{eq1ffn}) and (\ref{eq2ffn}) according to three unknown constants $f^F,
f^D, f^S$ one comes to the following expressions

\begin{eqnarray}
&&f_1^F=\frac{1}{2}\left[ \sqrt{3} \left(f_{\phi NN}^{(1)}\cos \theta-f_{\omega NN}^{(1)}\sin \theta\right) + f_{\rho NN}^{(1)}\right] \label{eqf66}\\\nonumber
&&f_1^D=\frac{1}{2}\left[3f_{\rho NN}^{(1)} -\sqrt{3}\left(f_{\phi NN}^{(1)} \cos \theta-f_{\omega NN}^{(1)}\sin \theta\right) \right] \nonumber \\ &&f_1^S=\sqrt{2}\left(f_{\omega NN}^{(1)}\cos \theta+f_{\phi NN}^{(1)}\sin
\theta\right) \nonumber
\end{eqnarray}

\begin{eqnarray}
&&f_2^F=\frac{1}{2}\left[ \sqrt{3} \left(f_{\phi NN}^{(2)}\cos \theta-f_{\omega NN}^{(2)}\sin \theta\right) + f_{\rho NN}^{(2)}\right] \label{eqf67}\\ \nonumber
&&f_2^D=\frac{1}{2}\left[3f_{\rho NN}^{(2)} -\sqrt{3}\left(f_{\phi NN}^{(2)} \cos \theta-f_{\omega NN}^{(2)}\sin \theta\right) \right] \nonumber \\ &&f_2^S=\sqrt{2}\left(f_{\omega NN}^{(2)}\cos \theta+f_{\phi NN}^{(2)}\sin
\theta\right) \nonumber
\end{eqnarray}

\begin{eqnarray}
&&f_1^{F'}=\frac{1}{2}\left[ \sqrt{3} \left(f_{\phi' NN}^{(1)}\cos \theta'- f_{\omega' NN}^{(1)}\sin \theta'\right) + f_{\rho' NN}^{(1)}\right]\label{eqf68} \\\nonumber &&f_1^{D'}=\frac{1}{2}\left[3f_{\rho' NN}^{(1)} -\sqrt{3}\left(f_{\phi' NN} ^{(1)}\cos \theta'-f_{\omega' NN}^{(1)}\sin \theta'\right) \right] \nonumber \\ &&f_1^{S'}=\sqrt{2}\left(f_{\omega' NN}^{(1)}\cos \theta'+f_{\phi' NN}^{(1)} \sin \theta'\right) \nonumber
\end{eqnarray}

\begin{eqnarray}
&&f_2^{F'}=\frac{1}{2}\left[ \sqrt{3} \left(f_{\phi' NN}^{(2)}\cos \theta'- f_{\omega' NN}^{(2)}\sin \theta'\right) + f_{\rho' NN}^{(2)}\right]\label{eqf69} \\\nonumber &&f_2^{D'}=\frac{1}{2}\left[3f_{\rho' NN}^{(2)} -\sqrt{3}\left(f_{\phi' NN} ^{(2)}\cos \theta'-f_{\omega' NN}^{(2)}\sin \theta'\right) \right] \nonumber \\ &&f_2^{S'}=\sqrt{2}\left(f_{\omega' NN}^{(2)}\cos \theta'+f_{\phi' NN}^{(2)} \sin \theta'\right) \nonumber
\end{eqnarray}
and by a substitution of numerical values (\ref{eqf59}), (\ref{eqf60}), (\ref{eqf64}), (\ref{eqf65}),
respectively, one finds the numerical values
\begin{eqnarray}
&&f_1^F=-5.69470  \\\label{eqf70}
&&f_1^D=9.4103  \nonumber \\
&&f_1^S=42.1706 \nonumber
\end{eqnarray}

\begin{eqnarray}
&&f_2^F=7.08952\\\label{eqf71}
&&f_2^D=21.6245 \nonumber \\
&&f_2^S=-7.1465 \nonumber
\end{eqnarray}

\begin{eqnarray}
&&f_1^{F'}=0.38580 \\\label{eqf72}
&&f_1^{D'}=20.45639 \nonumber \\
&&f_1^{S'}=-5.0842 \nonumber
\end{eqnarray}

\begin{eqnarray}
&&f_2^{F'}=6.0577 \\\label{eqf73}
&&f_2^{D'}=-41.7801 \nonumber \\
&&f_2^{S'}=-31.3390 \nonumber
\end{eqnarray}
of all scrutinized coupling constants in SU(3) invariant
interaction Lagrangians of vector-meson nonets with $1/2^+$ octet
baryons, which allow us to predict behaviors of EM FFs of all
$1/2^+$ octet hyperons in space-like and time-like regions,
including their real and imaginary parts, phase differences of
electric and magnetic FFs and corresponding differential and total
cross-sections, in which EM structure of hyperons is planned to be
measured. Predictions on all above mentioned quantities will be
given in the next paper which will be published elsewhere.

\section{Conclusions}
All existing 11 sets of data on nucleon EM FFs in space-like and time-like region from more than 40 different experiments has been analysed, from which
534 reliable experimental points are here reasonably described by 9 resonance $U\&A$ model of nucleon EM structure, which respects the SU(3) symmetry and OZI rule violation. The analysis revealed the following results:
\begin{itemize}
\item
an existence of the zero of $G_{Ep}(t)$ around $t_z=13\; \rm{GeV}^2$ is
again confirmed
\item
the value of the proton charge \textit{rms} radius coincides with
the value obtained in the muon hydrogen atom spectroscopy
experiment and in this way the existing puzzle is definitively
removed
\item
the value of the neutron charge mean squared radius is identical with the value given by Rev. of Part. Physics
\item
the coupling constants in the SU(3) invariant Lagrangian of the vector meson nonet interaction with $1/2^+$ octet baryons $f^F, f^D, f^S$ are determined numerically, which allow to predict all quantities describing the EM structure of $1/2^+$ octet hyperons.
\end{itemize}

~~~~~~~~~~

The work was supported by Slovak Grant Agency for Science VEGA, gr. No.
1/0158/13 and by Slovak Research and Development Agency APVV, gr. No.
APVV-0163-12.

\end{document}